\documentclass[
  journal=largetwo,
  manuscript=article-type,
  year=2025,
  volume=37,
]{cup-journal}

\usepackage{amsmath}
\usepackage[nopatch]{microtype}
\usepackage{booktabs}

\usepackage{comment}
\usepackage{graphicx}	
\usepackage{amsmath}	
\usepackage{amssymb}	

\usepackage{mathptmx}
\usepackage{txfonts}
\usepackage[T1]{fontenc}
\usepackage{hyperref}

\usepackage{xcolor}
\usepackage[normalem]{ulem} 

\title{HINORA II: Testing the Existence of the Council of Giants in $\rm \Lambda$CDM simulations}

\author{Edward Olex}
\affiliation{Departamento de F\'isica Te\'{o}rica, M\'{o}dulo 15, Facultad de Ciencias, Universidad Aut\'{o}noma de Madrid, 28049 Madrid, Spain}
\email[E. Olex]{edward.olex@uam.es}

\author{Alexander Knebe}
\affiliation{Departamento de F\'isica Te\'{o}rica, M\'{o}dulo 15, Facultad de Ciencias, Universidad Aut\'{o}noma de Madrid, 28049 Madrid, Spain}
\alsoaffiliation{Centro de Investigaci\'{o}n Avanzada en F\'isica Fundamental (CIAFF), Facultad de Ciencias, Universidad Aut\'{o}noma de Madrid, 28049 Madrid, Spain}
\alsoaffiliation{International Centre for Radio Astronomy Research, University of Western Australia, 35 Stirling Highway, Crawley, Western Australia 6009, Australia}

\author{Noam I. Libeskind}
\affiliation{Leibniz-Institut f\"ur Astrophysik Potsdam (AIP), An der Sternwarte 16, D-14482 Potsdam, Germany}

\author{Stefan Gottl\"ober}
\affiliation{Leibniz-Institut f\"ur Astrophysik Potsdam (AIP), An der Sternwarte 16, D-14482 Potsdam, Germany}

\author{Dmitry I. Makarov}
\affiliation{Special Astrophysical Observatory, Russian Academy of Sciences, Nizhnii Arkhyz, 369167 Russia}

\keywords{galaxies: haloes, Local Group, cosmology: theory, dark matter, large-scale structure} 

\begin{document}

\begin{abstract}

The discovery of the galaxy ring known as the Council of Giants (CoG) highlights the need to explain such structures in the Local Universe. In the first paper of this series we presented HINORA -- a code to locate (ring-like) structures in 3D point sets -- and used it to identify the CoG in the most complete observations of the Local Volume. Here, in Part~II, we apply the same method to cosmological simulations to quantify the possible existence of such objects in the $\rm \Lambda$CDM model of structure formation. We analyze DM-only simulations with random and constrained initial conditions, selecting regions that reproduce the properties of the Local Group and Volume, respectively. In order to use the same selection criteria as previsouly done for observations, we relate K-band luminosities to halo masses through semi-empirical relations. After confirming that the selected regions from the simulations match the observed mass function and density of the Local Universe, we use HINORA to search for ring-like structures in them. We find that the existence of CoGs in $\rm \Lambda$CDM simulations is a rather unusual phenomenon. The observed CoG represents an anomaly of more than 2.7$\sigma$ from what is expected in the distribution of massive galaxies in $\rm \Lambda$CDM. These results hint that the CoG could either be a rare chance configuration or the imprint of physical processes at intermediate scales that standard DM-only simulations fail to capture.

\end{abstract}
\section{Introduction}

The current era of astronomical surveys is revolutionizing our understanding of the Universe, revealing unprecedented details on both large and small scales. While redshift catalogs capture cosmic structures on vast scales, precise distance measures shed new light on the galaxies closest to the Milky Way (MW). In recent years, detailed analyses of distance and peculiar velocity catalogs in the Local Universe have uncovered hidden patterns in the distribution of galaxies, with important implications for modern cosmology. On small scales, attention has been drawn to the observation of kinematically stable satellite planes around the MW, Andromeda and also Centaurus A \citep[see][]{Bullock2017,Pawlowski2018, Tully2015,Muller2018}, which have been extensively studied in cosmological simulations \citep{Libeskind2005,Libeskind2009,Lovell2011,Wang2013,Cautun2015}, suggesting that their origin does not pose a major tension with the standard cosmological model \citep{Buck2015,Gillet2015,Ahmed2017,Shao2019,Samuel2021,Pham2023,Matias2024}.  

Flat structures are not limited to dwarf satellites. The massive galaxies within $\sim 10$ Mpc of the Milky Way, located in the so-called Local Volume (LV), form a flat arrangement \citep{Rubin1951,Vaucouleurs1953,Vaucouleurs1958} known as the Local Sheet \citep{Tully2008}. Recently, \citet{Peebles2023} pointed out that extensive sheet-like structures appear more frequently than expected, highlighting the lack of a clear explanation within $\rm \Lambda$CDM. Additional evidence suggests that Local Sheet-like structures are rarely reproduced in cosmological simulations \citep{Neuzil2020,Calvo2023}, although constrained simulations appear more capable of generating such configurations \citep{Sawala2024}.

Our research focuses on studying these types of flat regular structures simultaneously in observations and simulations, with particular emphasis on one case: the ring of massive galaxies known as the Council of Giants (CoG), discovered by \citet{McCall2014} in his dataset. In that paper, McCall argues that the CoG consists of 12 massive galaxies (10 of them spirals) forming a 3.75 Mpc ring-like structure embedded in the Local Sheet and containing the Local Group (LG) near its center.  

In our first paper (\cite{Olex2024}, from now on referred to as Paper \hyperlink{part1}{I}), we developed High-Noise RANSAC (HINORA), a method that enables the identification of regular patterns in galaxy catalogs from both observations and simulations. By applying HINORA to the updated Local Volume Galaxies survey \citep[LVG;][]{Karachentsev2013}, in Paper \hyperlink{part1}{I} we located the CoG in the regime of massive galaxies, as well as a second, similar ring within the LV when all luminosity regimes were considered. We also detected both patterns statistically by analyzing the clustering properties of the survey.  

The existence of these galaxy rings in the LV could provide valuable insight into how the Local Sheet and the LG were formed, and perhaps even reveal new, still-unknown phenomena in the general process of structure formation at intermediate scales. However, their existence might also be the result of purely coincidental alignments created by random motions of nearby galaxies. This raises a fundamental question: are galaxy rings like the CoG signatures of new physics, or simply chance alignments? The present, second paper aims to shed light on this issue.  

For this purpose, we have selected in cosmological dark matter (DM) only simulations environments analogous to that surrounding the LG. We extract the simulated LVs from two main types of simulations: those with random initial conditions and those with constrained phases - such as the HESTIA \citep{Libeskind2020} produced by the CLUES collaboration \footnote{https://www.clues-project.org/} - specifically designed to reproduce the cosmography near the LG. Both sets satisfy similarity criteria defined according to the known properties of the LV. Using halo luminosity-mass relations, we restrict the simulated objects to the massive regime where the observed CoG resides, and we apply the same HINORA analysis as was performed for the LVG. This enables us to quantify how often rings appear in the simulated LVs and to estimate the probability of their formation within $\rm \Lambda$CDM.  

This paper is structured as follows. Section~\ref{sec:LV} describes the cosmological simulations and our procedure to identify LV analogues in them. Section~\ref{sec:method} summarises the observed properties of the CoG from Paper~\hyperlink{part1}{I} and the simulation-to-observational mapping adopted by us. In Section~\ref{sec:LVprop} we compare the simulated LV realisations with the observed LVG catalogue. Section~\ref{sec:HINORA} presents the application of HINORA to search for CoG-like rings in those environments. We close with a brief discussion and conclusions in Section~\ref{sec:conc}.

\section{Local Volume simulations}
\label{sec:LV}

The LV is, so far, the only region of the Universe in which a $\sim 4$ Mpc-radius ring of massive galaxies has been found. In Paper \hyperlink{part1}{I}, we used the LVG catalog to study this region of the Universe, which consists of 1069 galaxies within 10 Mpc of the MW. In this section, we present the different types of simulations we use to reproduce this environment and study the potential formation of CoGs in it.

\subsection{Constrained simulations}

To reproduce as closely as possible the observed cosmography of the LV, we use the HESTIA cosmological simulations \citep{Libeskind2020} that are based upon constrained initial conditions. Simulations such as HESTIA and others carried out by the CLUES collaboration \citep{Gottloeber2010,Libeskind2010,Carlesi2016,Sorce2016} successfully generate environments that resemble both the LG \citep{Libeskind2020} and the surrounding LV \citep{Wempe2025}, by reconstructing the large-scale structure of the Local Universe. To do this, CLUES simulations use observables that allow the local density field to be reconstructed through a Wiener Filter, thereby obtaining constraints on the Gaussian field that generates the initial conditions \citep{Hoffman1991,Zaroubi1995}. See \citet{Hoffman2009} for a more complete summary of this technique. In the case of HESTIA, the constrained initial conditions are based on the CosmicFlows-2 survey \citep[CF2,][]{Tully2013} using the observed galaxy positions and peculiar velocities. 

Since these simulations do not constrain small scales (i.e. non-linear scales corresponding to the LG and below), several exploratory runs are required to reproduce the finer details of the LG. In practice, nearly $10^3$ low-resolution constrained initial phases were first generated in periodic boxes of side $100\,h^{-1}$ Mpc ($147.5$ Mpc assuming \citet{PlanckXVI} based cosmology), each containing $256^{3}$ DM particles. For each run, a Lagrangian region of radius $\simeq 14.7$\,Mpc ($10\,h^{-1}$\,Mpc) that collapses to the LV at $z=0$ was identified and re-simulated at high resolution following the prescription of \citet{Katz1993}, replacing the original particles with an effective resolution of $512^{3}$ within the central sphere. This resolution is sufficient to resolve LG-like halos with several thousand particles.

The particular set used here consists of DM-only simulations of a ``zoomed'' sphere with radius $10\,h^{-1}$\,Mpc ($\simeq 14.7$\,Mpc) and a mass resolution of $m_{\rm DM}=6\times10^{8}\,M_{\odot}$, embedded in a periodic box of $100\,h^{-1}$\,Mpc. They were evolved with \textsc{Arepo} \citep{Weinberger2020} using the \citet{PlanckXVI} based cosmology $\Omega_{m}=0.318$, $\Omega_{b}=0.048$, $\Omega_{\Lambda}=0.682$, $\sigma_{8}=0.83$, $n_{s}=0.96$, and $h=0.677$. Halos are identified with the AMIGA Halo Finder \citep[AHF,][]{Knollmann2009}, using the $M_{200}$ definition, i.e. the mass within a sphere of 200 times the critical density.

For the present work we start with selecting only those realisations that contain a pair of halos that satisfy the main properties of the LG based on the observed properties of the MW-M31 pair \citep[see Section~3.2 of][]{Libeskind2020}. 

We will refer to these properties as \textbf{HESTIA criteria} , and they are as follows:

\begin{itemize}
    \item[a)] Two haloes of mass: $8 \times 10^{11} < M_{200}/M_{\odot} < 3 \times 10^{12}$.
    \item[b)] Separation: $0.5 < d_{\rm sep}/{\rm Mpc} < 1.2$.
    \item[c)] Isolation: no third halo more massive than the smaller one within 2 Mpc of the midpoint.
    \item[d)] Halo mass ratio of smaller to larger halo $> 0.5$.
    \item[e)] Infalling, i.e. $v_{\rm rad} < 0$.
\end{itemize}

The numerical values adopted for the HESTIA criteria are not meant to impose tight constraints on the detailed properties of the LG, but rather to define a broad observationally motivated window that selects plausible LG analogues. The adopted mass window reflects current observational estimates that place MW at about $1.0-2.1 \times10^{12}\,M_{\odot}$  \citep{Posti2019, Hattori2018, Monari2018, Watkins2019, Zaritsky2017} and M31 at about $0.6-2.0 \times10^{12}\,M_{\odot}$ \citep{Kafle2018, Tamm2012, Diaz2014, Corbelli2010}. Similarly, the separation and kinematic ranges are chosen to encompass the current observational uncertainties on the MW-M31 system \citep[see the discussion in Section~3.2 of][]{Libeskind2020}. 

If the HESTIA criteria are met, the LG-like pair must be found within 5 Mpc of its expected position in the simulation, namely at the supergalactic origin: (SGX, SGY, SGZ) = (0, 0, 0). No other cluster more massive than the simulated Virgo is allowed within a sphere of 20 Mpc centered on the LG\footnote{We also require a Virgo-like cluster (with a mass $> 2 \times 10^{14} M_{\odot}$) to be present, meaning that the constrained simulations must produce a cluster at the Virgo position. This is a relatively ``light'' criterion, as the constrained initial conditions algorithm nearly always forms a Virgo cluster.}. We note that the cluster selection has two complementary components. In all LV-like samples we impose a baseline requirement that no halo with $M_{200}>2\times10^{13}\,M_{\odot}$ is present within the 10\,Mpc LV, in order to exclude strongly cluster-dominated environments. In addition, the HESTIA constrained realizations are required to host a unique Virgo-like cluster at the observed CF2 position \citep[$\sim16$ Mpc from MW, see][]{Libeskind2020}, reflecting the goal of reproducing the observed large-scale cosmography.
After applying these criteria, a total of \textbf{64} different realizations of the LG are obtained, distinguished by the random-phase seeds that set their small-scale structure. As in our first work with the LVG survey, we restrict our analysis to galaxies within 10 Mpc of the candidate MW in each HESTIA simulation.

\subsection{Random simulations}

Given the success of the HESTIA simulations in reproducing various features of the Local Universe beyond the MW - Andromeda pair, we also adopt the same criteria to find LVs in random-type (i.e. unconstrained) simulations. To create a contrasting sample representative of a generic, unconstrained universe, we apply the five LG selection criteria to the Small MultiDark Planck simulation \citep[SMD,][]{Klypin2016}.
The SMD simulation is a $400\,h^{-1}$\,Mpc periodic box with a particle mass resolution of $m_{MD} = 9.6\times10^{7}\,M_{\odot}$, evolved with \textsc{Gadget-2} \citep{Springel2005} under a \citet{PlanckXVI} based cosmology of $\Omega_{m}=0.307$, $\Omega_{b}=0.048$, $\Omega_{\Lambda}=0.693$, $\sigma_{8}=0.83$, $n_{s}=0.96$, and $h=0.678$. Halos are identified with the ROCKSTAR (Robust Overdensity Calculation using K-Space Topologically Adaptive Refinement) halo finder \citep{Rockstar2013}, using the $M_{200}$ definition. 

Once we locate LG-like galaxy pairs in SMD that satisfy all five criteria and the baseline cluster exclusion (i.e., no halos with $M_{200}>2\times10^{13}\,M_{\odot}$ within $10$ Mpc), we define their LVs as catalogs of all galaxies within $10$\,Mpc of the MW candidate (the less massive member of the LG pair). This procedure yields 4430 candidate volumes; after discarding those overlapping by more than 50\% of their volume, we obtain a final set of \textbf{4048} independent LVs.

To test whether ring-like structures arise directly from intrinsic LV properties, we generate a third reference sample by placing $10^4$ random spheres of radius $10$\,Mpc within the SMD box without any selection criteria, a number large enough to provide a statistically representative sampling of the simulation volume. After excluding regions overlapping by more than 50\%, this yields \textbf{9292} independent random LVs.  
Table~\ref{tab:lv} summarizes the three sets of simulated spherical volumes used in this work to search for possible galaxy rings.

\begin{table}
 \caption{List of the simulated volumes in which we have searched for the Council of Giants using HINORA. Each volume consists of a sphere of radius 10 Mpc by similarity to LVG.}
 \label{tab:lv}
 \begin{center}
 \begin{tabular}{>{\centering\arraybackslash}m{2cm} >{\centering\arraybackslash}m{1.7cm} >{\centering\arraybackslash}m{1.7cm} >{\centering\arraybackslash}m{1.7cm}}
 \hline\hline
 &HESTIA&SMD&SMD Random\\
 \hline
 \noalign{\vspace {.1cm}}
 Number&64&4048&9292\\
 \noalign{\vspace {.1cm}}
 Simulation&HESTIA&Small MultiDark Planck&Small MultiDark Planck\\
 \noalign{\vspace {.1cm}}
 Initial conditions&Constrained&Random&Random\\
 \noalign{\vspace {.1cm}}
 Local Group&HESTIA criteria&HESTIA criteria&-\\
 \noalign{\vspace {.1cm}}
 Exclude cluster environments&No LV halo with $M_{200} > 2\times10^{13}M_{\odot}$&No LV halo with $M_{200} > 2\times10^{13}M_{\odot}$&-\\
 \noalign{\vspace {.1cm}}
 Virgo-like cluster at the CF2 position&Yes&-&-\\
 \noalign{\vspace {.1cm}}
 \hline\hline
 \end{tabular}
 \end{center}
 \end{table}

All simulations employed here are DM-only and we restrict our analysis to masses well above the resolution limits. Two different halo finders are used: AHF for HESTIA and ROCKSTAR for SMD. The slight differences in $\Omega_m$ and $h$ between HESTIA and SMD are likewise negligible for the scale studied.

\section{Methodology}
\label{sec:method}

In this section we describe the approach used to search for CoG-like structures within our set of $\rm \Lambda$CDM simulations.  
We first recall the observational definition of the CoG established in Paper~\hyperlink{part1}{I} and describe how the HINORA algorithm identifies such ring-like systems.  
Next, we explain how observational quantities (K-band luminosities) are converted into DM halo properties to facilitate the comparison with simulated LVs.  

\subsection{The detection of the Council of Giants}
\label{sec:CoG}

We focus on the CoG-like ring identified in Paper \hyperlink{part1}{I}, found using HINORA in the regime of more massive galaxies, with luminosity cuts of $\mathrm{log}_{10}(L_{K}) > 9, 10, 10.5$, where $L_K$ is the luminosity in the $K$-band. This structure, formed by the brightest galaxies in the LV, shows consistent properties across all cuts and exhibits the same geometrical features as the CoG reported by \citet{McCall2014} (see Figs.~5 and 6 in Paper \hyperlink{part1}{I}). 

The HINORA method used to identify the CoG in the LVG is based on the RANdom SAmple Consensus (RANSAC) algorithm, which applies a non-deterministic strategy to detect regular structures in point clouds. To find a ring-like pattern in the data, RANSAC randomly selects three points and defines the circle that passes through them, recording all points located within a distance $\tau$ of that circle. The configuration that includes the largest number of points is then considered the best-fitting hypothesis. However, this approach has a key limitation: since it relies solely on maximizing the number of points, it will always identify a single structure even in a purely random distribution.

HINORA modifies RANSAC to overcome this issue by quantifying three properties of any detected structure using three new parameters: 

\begin{itemize}

\item[i)] $\alpha$, to quantify the noise in the data. This parameter is defined as the ratio $\alpha = N_I/(N_I + N_O)$ between the number of inliers $N_I$ and outliers $N_O$ associated with a given model. Its local value is compared with the global prediction $\bar{\alpha}$, which represents the expected value of $\alpha$ for the full dataset.

\item[ii)] $\beta$, to quantify the regularity of the data. If $\phi_i$ denotes the angular separations between consecutive points along the detected ring, this parameter is defined as the ratio between the standard deviation $\sigma_{\phi}$ and the mean value $\langle \phi \rangle$ of these angles, $\beta = \sigma_{\phi}/\langle \phi \rangle$. Its value is compared with the global expectation $\bar{\beta}$ derived for the full dataset.

\item[iii)] $n_I$, to quantify the overall statistical significance of the structure. It is defined as the fraction of inliers,
$n_I = N_I / N_{\mathrm{tot}}$,
where $N_{\mathrm{tot}}$ is the total number of data points. This value is compared with the expected inlier fraction $\bar{n}_I$, which is fixed a priori.

\end{itemize}

Once the expected values of $\bar{\alpha}$ and $\bar{\beta}$ are computed analytically for a given dataset, and $\bar{n}_I$ is specified, only structures satisfying $\alpha < \bar{\alpha}$, $\beta < \bar{\beta}$, and $n_I > \bar{n}_I$ are accepted. For a more detailed description of HINORA as well as the analytical prediction of $\bar{\alpha}$ and $\bar{\beta}$, see the Section 3 of Paper~\hyperlink{part1}{I}.

HINORA depends on three parameters, which are set a priori rather than inferred from the data. In Paper~\hyperlink{part1}{I}, these were defined as follows:
\begin{itemize}
    \item[a)] $\tau$, the allowed thickness of the ring. We adopt an inner radius of $\tau = 1$ Mpc.
    \item[b)] The minimum fraction of galaxies or halos that two detected rings must share in order to be considered the same structure. This prevents the method from identifying multiple slightly displaced copies of the same ring. We use a value of 30$\%$.
    \item[c)] $\bar{n_I}$, as mentioned above the minimum fraction of the total galaxies or halos that a ring must contain to be accepted. We use $\bar{n_I} = 0.15, 0.20,$ and $0.25$.
\end{itemize}

In this second paper, we keep the first two parameters fixed to the values adopted in Paper \hyperlink{part1}{I}, where they were shown to have no significant impact on the results, and vary only the third. The value of $\bar{n_I}$ can strongly influence the identification of structures: low values may introduce spurious detections, while high values can suppress genuine ones (see Section~4 of Paper \hyperlink{part1}{I}). All other ring properties (e.g., radius, position, or orientation) are not imposed by the method but are directly inferred from the data. This is a key advantage of HINORA, as it avoids introducing biases in the geometry of the detected structures. 

Table~\ref{tab:cog} lists the characteristics of the observed CoG obtained by applying HINORA to the unmodified LVG catalog.\footnote{In Paper \hyperlink{part1}{I}, we modified the observational data to estimate the effect of distance uncertainties; see Section~4 of that paper for details.} The geometrical parameters of the CoG remain remarkably stable across different luminosity cuts. We quantify this stability by measuring the maximum variation in three independent properties of the rings: the position of the center (in Mpc), the orientation angle between their planes (in degrees), and the radius (in Mpc), denoted as $\Delta_{\rm c}$, $\Delta_{\rm a}$, and $\Delta_{\rm r}$, respectively.

\begin{table}

 \caption{Characteristics of the Council of Giants detected by HINORA in Paper \protect\hyperlink{part1}{I}. The ``SG'' coordinates are in the Supergalactic reference system, while the orientation is normalized to 1.}

 \label{tab:cog}
 \begin{center}
  \begin{tabular}{>{\centering\arraybackslash}m{1.5cm} >{\centering\arraybackslash}m{1cm} >{\centering\arraybackslash}m{1cm} >{\centering\arraybackslash}m{1cm}>{\centering\arraybackslash}m{1cm} >{\centering\arraybackslash}m{0.5cm}}
 \hline\hline
 & Coord.&Cut 1&Cut 2&Cut 3& $\Delta$\\
 \hline
 \noalign{\vspace {.1cm}}
 $\mathrm{log}(L_k) >$&&9&10&10.5&\\
 \noalign{\vspace {.0cm}}
 $M_{200}/10^{11} M_{\odot} > $ && 1.01&2.84&6.06&\\
 \noalign{\vspace {.3cm}}
 &SGX& -1.11&-0.35&-0.52&\\
 Center (Mpc)&SGY&0.73&0.12&0.68&1.10\\
 &SGZ&0.28&-0.22&-0.22&\\
 \noalign{\vspace {.3cm}}
 &X&0.06&-0.06&-0.06&\\
 Orientation&Y&0.01&-0.04&-0.02&7.24$^{\circ}$\\
 &Z&0.99&0.99&0.99&\\
 \noalign{\vspace {.3cm}}
 Radius (Mpc) &&4.35&3.83&3.76&0.59\\
 \noalign{\vspace {.3cm}}
 Galaxies &&26&10&9&\\
 \noalign{\vspace {.3cm}}
 $n_{I}$ &&0.22&0.22&0.29&\\
\noalign{\vspace {.1cm}}
 \hline\hline
 \end{tabular}
 \end{center}
 \end{table} 

These quantities express the maximum variation of the CoG geometry across luminosity cuts, and are used to identify as a single object a structure that appears in several mass-selected samples with slightly different positions, orientations, and sizes. For each geometric property, the $\Delta$ value corresponds to the maximum pairwise difference measured between the CoGs detected at different luminosity cuts. As shown in Table~\ref{tab:cog}, the maximum displacement of the CoG center is $\Delta_{\rm c} = 1.10$ Mpc, computed as the largest three-dimensional separation between any two centers listed in the table. Similarly, the maximum variation in orientation, $\Delta_{\rm a} = 7.24^{\circ}$, corresponds to the largest angle between the planes defining the rings at different cuts, and the maximum change in radius is $\Delta_{\rm r} = 0.59$ Mpc. 

We adopt these maximum variations as tolerance thresholds when defining a CoG in the simulations, requiring that candidate rings exhibit the same or smaller level of geometric persistence (in all three parameters) across different mass cuts.

\subsection{Connecting simulations with observations}

On the one hand we are dealing with observed properties like the K-band luminosity of galaxies, but on the other hand the simulations provide halo masses. In order to apply comparable cuts to the object properties, we convert the K-band luminosities used in Paper \hyperlink{part1}{I} into stellar masses ($M_{\star}$), and then relate those stellar masses to the halo masses ($M_{200}$) using semi-empirical relations. Infrared bands are best suited for this purpose because their $M_{\star}/L$ relation is less sensitive to variations in stellar populations, age, star-formation history (SFH) and metallicity than in the optical bands \citep{Rock2015,Wen2013,Meidt2014}. In particular, the K-band provided by LVG is commonly used to estimate stellar mass: it suffers less dust extinction and is less affected by SFH \citep{Bell2003,Brinchmann2000,Cole2001,Bundy2005,Taylor2011,Beare2019}. We therefore use the K-band to estimate $M_{\star}$ for LVG objects using several published relations, and subsequently derive their corresponding $M_{200}$ mass. When an object lacks direct K-band data, we will adopt the K-band estimates provided by \citet{Karachentsev2013}, who derive them from other measured bands as described in their paper.

The empirical relation we use for the inference of the stellar mass is:

\begin{equation}
M_{\star}/L_{K} \sim 0.6 M_{\odot}/L_{\odot}
\label{eq:Lelli}
\end{equation}

demonstrated by \citet{McGaugh2014}, and \citet{Lelli2016}; self-consistent with inference in other bands. None of the galaxies studied deviates more than $\sigma_{K \rightarrow \star} \sim 0.1$ dex from this color-independent function, which corrects the overestimation of stellar mass in K-band from previous constant relations by \citet{Bell2003} or \citet{Into2013} \citep{Beare2019}.

Once $M_{\star}$ is estimated, we apply the semi-empirical stellar-halo mass relation (SHMR) of \citet{Puebla2017} via abundance matching, with a scatter of less than $\sigma_{\star \rightarrow h} \sim 0.15$ dex. Specifically, we interpolate the tabulated numerical results for the mean inverted SHMR at $z=0$ to determine the mean virial mass, $M_{h}$, for a given $M_{\star}$.

Since the SHMR contains the virial mass of the halo, we adopt
\begin{equation}
M_h \approx 1.22\,M_{200}
\label{eq:conversion}
\end{equation}
to convert between mass definitions, consistent with typical NFW concentration parameters for galaxies in this mass range \citep{White2001,Pierpaoli2003}. Therefore we adopt an average conversion and account for the resulting uncertainty induced by the intrinsic scatter in the mass–concentration relation and by deviations from ideal NFW profiles. Recent studies that explicitly propagate these effects find a typical scatter of $\sigma_{h\rightarrow200} \sim 0.1$ dex for mass conversions that do not assume an individual concentration value \citep{Ragagnin2021,Diemer2015,Dutton2014}. 

This procedure yields estimates of the $M_{200}$ values corresponding to the $L_K$ luminosity cuts that define the CoG, with further details provided in the \ref{sec:app1}. The total scatter from the relation between K-band luminosity and $M_{200}$ is  

 \begin{equation}
    \sigma_{K\rightarrow200}^2 \approx \sigma_{K \rightarrow \star}^2 + \sigma_{\star \rightarrow h}^2 + \sigma_{h\rightarrow200}^2, 
\end{equation}
which results in $\sigma_{K\rightarrow200} \sim 0.2$. 

\section{Local Volume properties}
\label{sec:LVprop}

\begin{figure}[h]
\centering
\includegraphics[width=\columnwidth]{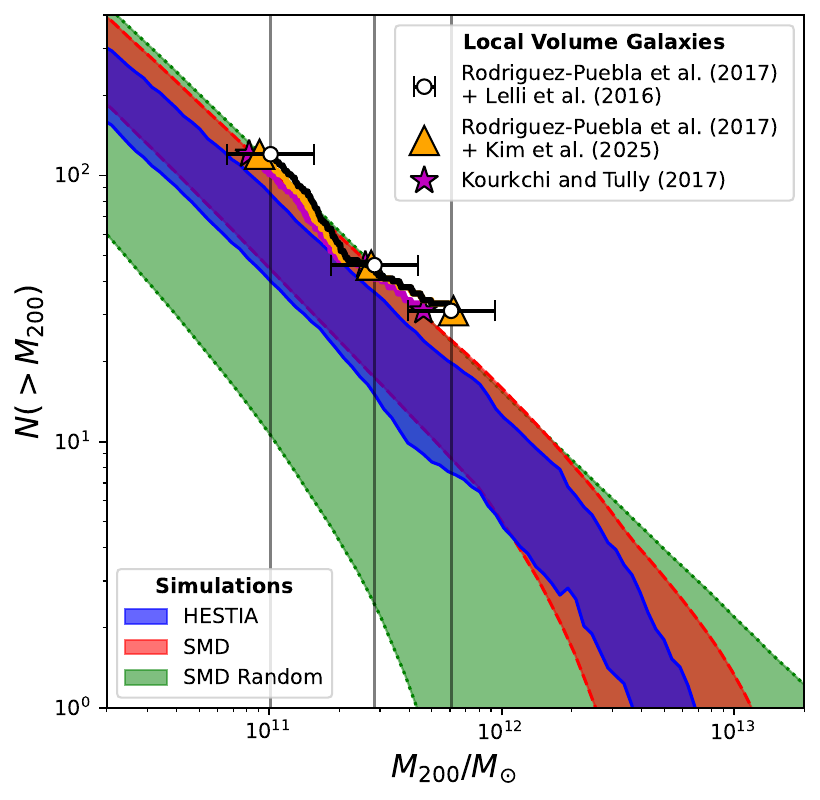}
\caption{Cumulative halo mass function for $M_{200}$ in each of the Local Volumes extracted from the HESTIA constrained simulations (blue region) and SMD random simulation (red region). Also shown in green are randomly placed volumes in SMD. The simulations are compared with those obtained for the LVG survey using three different $M_h / L_k$ relations. The vertical lines indicate the main mass cuts at which HINORA was applied in Sec.~\ref{sec:HINORA}.}
\label{fig:HMF}
\end{figure}

The $M_{200}$ estimates enable a direct comparison between the DM halos in the simulations and the galaxies observed in the LVG. To examine possible differences in their mass distributions, Fig.~\ref{fig:HMF} shows the cumulative halo mass function (CHMF) for all LVs identified in the HESTIA (solid blue line) and SMD (dashed red line) simulations. The shaded regions indicate the $\pm 1\sigma$ scatter around the mean. Randomly placed spherical volumes within the SMD box (dotted green line) are also shown for reference. As these random volumes are unconstrained, they exhibit a much larger intrinsic scatter. Both HESTIA and SMD LVs display systematically higher halo abundances than random volumes, as expected since the LG forms in a relatively dense region of the cosmic web \citep{ForeroRomero2015}. Vertical lines indicate the $M_{200}$ thresholds corresponding to the luminosity cuts where the CoG was identified in Paper~\hyperlink{part1}{I} (see Table~\ref{tab:cog}).

Figure~\ref{fig:HMF} also includes the observed halo abundance in the LVG, derived using the \citet{Lelli2016} and \citet{Puebla2017} relations (L+RP) discussed earlier with the total scatter represented by the error bars. For comparison, two additional conversions between $L_K$ and $M_{200}$ were tested. The first combines \citet{Kim2025} with \citet{Puebla2017} (K+RP), using synthetic spectral energy distributions to estimate $M_{\star}/L_K$. The second, from \citet{Kourkchi2017} (KT), provides a direct relation between $\mathrm{log}(M_h/L_K)$ and has been applied previously to LVG analyses of satellite kinematics \citep{Karachentsev2021}. Details of these conversions and their implementation are given in \ref{sec:app1}. We find that the systematic differences among the three relations (L+RP, K+RP, and KT) are consistently smaller than the intrinsic scatter of L+RP, which we therefore adopt as our fiducial relation and reference uncertainty. The most notable result is the systematic excess of massive halos in the LVG compared to the simulated LV-like environments, particularly in the CoG mass regime. A similar excess relative to $\Lambda$CDM expectations has been reported previously using B-band luminosities in the LVG \citep{Neuzil2020}. 
\\ 

Another key property that characterizes the LV environment is its matter density. To assess this, we compute for each LV the halo density contrast relative to the cosmic mean. We define $\rho_h$ as the sum of the masses of halos with $M_{200}>10^{11}\,M_{\odot}$ (or equivalently $L_K>10^{9}\,L_{\odot}$) divided by the LV volume, $V=(4/3)\pi (10\,\mathrm{Mpc})^3$:

\begin{equation}
    \rho_h = \sum_{i}\frac{M_{200}^i}{V} ; \; \; \; \; M_{200}^i>10^{11}M_{\odot}
\label{eq:nh}
\end{equation}

The mean background halo density, $\bar{\rho}_h$, is obtained by summing all halos with $M_{200}>10^{11}\,M_{\odot}$ in the full SMD simulation and dividing by its total volume ($590\,\mathrm{Mpc}$)$^3$. The halo density contrast for each LV is then

\begin{equation}
    \delta_h = \frac{\rho_h - \bar{\rho_h}}{\bar{\rho_h}}
\label{eq:deltah}
\end{equation}

which, as in the standard definition, distinguishes overdense regions ($\delta_h>0$) from underdense ones ($\delta_h<0$).

\begin{figure}
\centering
\includegraphics[width=\columnwidth]{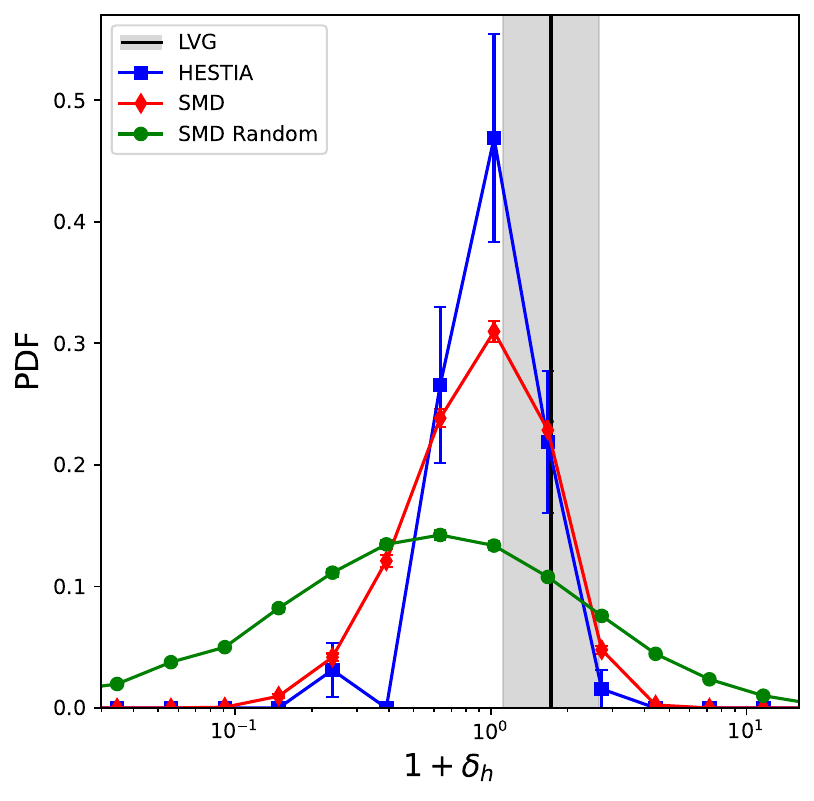}
\caption{Probability distribution of the halo density contrast (defined in equation \ref{eq:deltah}) for different Local Volumes. The different colors show the LVs obtained in HESTIA following all criteria (blue), SMD with all criteria (red), and SMD without any criteria (green). Value calculated for LVG survey and its associated scatter are shown with the black line + gray region.}
\label{fig:PDF}
\end{figure}

Figure~\ref{fig:PDF} shows the probability distribution function (PDF) of $1+\delta_h$ for the LVs identified in HESTIA and SMD. Observational results from LVG using the L+RP relation are also included for comparison (black line + gray area). The randomly placed spherical volumes exhibit large variations in total mass, ranging from nearly empty voids to dense clusters within the SMD box. In contrast, the constrained HESTIA and LG-like SMD volumes tend to occupy intermediate-density environments, highlighting the particular cosmographic conditions of the LG. A mild excess in halo mass within the LVG, already apparent in the CHMF, is again visible in this distribution.

These results indicate that the simulated LVs reproduce environments broadly comparable to that of the LVG, showing only a mild excess of mass, consistent with previous findings for the LG in cosmological simulations. The next step is to investigate whether CoG-like structures can naturally emerge within such environments, and, if they do, to determine how frequently they occur.

\section{The Council of Giants in $\rm \Lambda$CDM}
\label{sec:HINORA}

When eventually applying HINORA to our suite of simulations, any ring accepted in a single mass cut is treated as a ``candidate''\footnote{Since we work in a regime of massive galaxies, some mass cuts (and some LVs) may contain very few objects, so small values of $\bar{n_I}$ can produce spurious detections with only a few halos. To reduce false positives, we discard any accepted candidate with fewer than 8 inliers.} and then enters a second filtering stage. In that stage we require a candidate to be detected in all three mass cuts, and to satisfy the persistence thresholds $\Delta_c$, $\Delta_a$, and $\Delta_r$ defined in Section~\ref{sec:CoG}. For the simulated volumes we adopt the same set of HINORA parameters used for the LVG catalog in Paper \hyperlink{part1}{I}, increasing the number of iterations $N$ by a factor of ten to improve convergence.\footnote{See equation (1) of Paper \hyperlink{part1}{I}.} 

\begin{figure}
\centering
\includegraphics[width=\columnwidth]{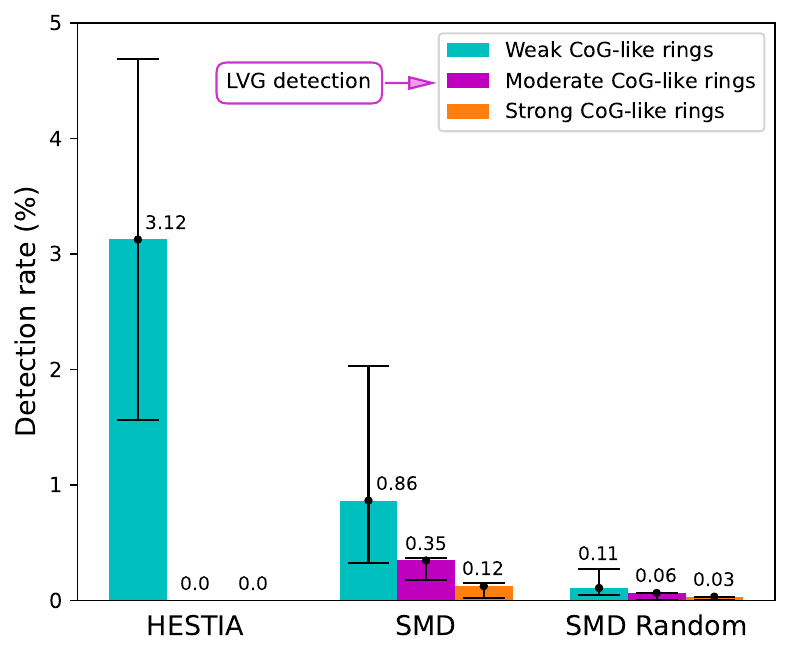}
\caption{Fraction of galaxy rings detected in the different sets of simulated Local Volumes.  
The statistical strength of each detection is classified as \textit{weak}, \textit{moderate}, or \textit{strong} according to the minimum fraction of halos in the volume that belong to the ring, requiring $n_I>0.15$, $n_I>0.20$, and $n_I>0.25$, respectively. The error bars consider the total uncertainty in the $L_{K} \rightarrow M_{200}$ relation.
}
\label{fig:Ring_frec}
\end{figure}

As laid out in Sec.~\ref{sec:CoG}, we also explore three values of the minimum halo fraction that define an accepted ring, $\bar{n_I}=0.15,\;0.20,\;0.25$, which we use to classify a statistical detection as \textit{weak}, \textit{moderate} or \textit{strong}, respectively. Note that the CoG is detected in the observations at $\bar{n_I}=0.20$ (moderate presence), since $n_I>0.20$ in all luminosity cuts (see Table~\ref{tab:cog}). After applying HINORA, the detection rate is computed as the number of volumes with a detected ring divided by the total number of volumes. This ratio is plotted in Fig.~\ref{fig:Ring_frec}, for the three choices of $\bar{n_I}$. While the bar represents the central value for mass cuts calculated with L+RP, the error bars are obtained by repeating the process for mass cuts shifted by the total scatter $\pm \sigma_{K\rightarrow200}$.

We did not find any LV with more than one ring in any sample. The main feature of Fig.~\ref{fig:Ring_frec} is the low detection rate (below $\sim5\%$ in all cases), which indicates that reproducing a structure with the geometry of the CoG is unlikely in the $\rm \Lambda$CDM simulations we tested. The most optimistic case (3.12\%) occurs in the HESTIA sample, suggesting that constrained initial phases slightly favour the appearance of CoG-like rings. However, this higher rate is attained only for weak rings; more significant (moderate or strong) structures are absent. Lower detection rates are found in the SMD sample and in the randomly placed volumes, the latter containing the fewest rings. The systematic difference between SMD and the random sample across all ring types points to an influence of the LG-selection criteria on the incidence of CoG-like structures. Moreover, because underdense regions occupy most of the simulated volume, the random spheres are predominantly located in such environments, which naturally explains their very low CoG detection rate.

\subsection{Rings in random vs constrained initial conditions}

Table~\ref{tab:hinora} gives the number of volumes of each type that contain a ring (without considering the scatter) under the condition $n_{I}>\bar{n_I}$. For $\bar{n_I}=0.15$, 35 confirmed rings are obtained in selected LVs of SMD. This implies that for a random draw of $64$ LVs from SMD, the mean number of LVs with ring is 0.55, so the weak-ring detection in HESTIA is about $2/0.55\approx 3.6$ times more favorable than in SMD with LG criteria, and 28.5 times more favorable than in random volumes. 

\begin{table}
\scriptsize
 \caption{Rings detected by HINORA in the three simulated samples used in this work, as well as the expected average number of rings $E[x]$ from 64 random draws from each sample.}
 \label{tab:hinora}
 \begin{center}
 \begin{tabular}{>{\centering\arraybackslash}m{1.5cm} >{\centering\arraybackslash}m{1cm} >{\centering\arraybackslash}m{1cm} >{\centering\arraybackslash}m{1cm} >{\centering\arraybackslash}m{1cm}}
 \hline\hline
  \noalign{\vspace {.1cm}}
 &&HESTIA&SMD&SMD Random\\
 \hline
 \noalign{\vspace {.1cm}}
 Number of volumes&&64&4048&9292\\
 \noalign{\vspace {.3cm}}
 &Weak $\bar{n_{I}} = 0.15$&2&35&10\\
 Rings found by HINORA& Moderate $\bar{n_{I}} = 0.20$&0&14&6\\
 &Strong $\bar{n_{I}} = 0.25$&0&5&3\\
\noalign{\vspace {.3cm}}
 &Weak $\bar{n_{I}} = 0.15$&2&0.55 $\pm$0.73&0.07 $\pm$0.26\\
 $E[x]$& Moderate $\bar{n_{I}} = 0.20$&0&0.22 $\pm$0.47&0.04 $\pm$0.20\\
 &Strong $\bar{n_{I}} = 0.25$&0&0.08 $\pm$0.28&0.02 $\pm$0.14\\

 \noalign{\vspace {.1cm}}
 \hline\hline
 \end{tabular}
 \end{center}
 \end{table}

To enable a proper comparison between simulations with constrained and random initial conditions without considering the scatter, we repeat the procedure of drawing 64 random SMD samples multiple times and computing the mean and standard deviation of the number of detected rings. This can be calculated analytically if the number of rings is treated as a random variable $x$ of a hypergeometric distribution with mean $E[x]$:

\begin{equation}
    E[x] =\frac{n_s k}{N}
\label{eq:meanh}
\end{equation}
where for comparison with HESTIA we set the number of samples $n_s=64$. $k$ is the number of LVs with ring found, while $N$ is the total number of LVs in each case. The variance of the distribution is:
\begin{equation}
    \sigma^2[x] = n_{s} \frac{N-n_{s}}{N-1} \frac{k}{N} \frac{N-k}{N} 
\label{eq:stdh}
\end{equation}

Table~\ref{tab:hinora} contains $E[x]$ for each LV type and each value of $\bar{n_I}$, with the error representing the standard deviation calculated with the equation (\ref{eq:stdh}). Given the small counts involved, Poisson (counting) uncertainties are large and these factors should be taken as indicative. Note that in less optimistic scenarios, i.e. $\bar{n_I}=0.20$ and $\bar{n_I}=0.25$, HESTIA contains no rings, which means that the probability has to be at least less than $100/64\approx1.56\%$. The hypergeometric mean of SMD in the $n_I=0.20$ case is $0.22$, so a null result in HESTIA is compatible with the order of magnitude in which it differs from SMD since $0.22 \times 3.6 < 1$.

\subsection{The significance of the Council of Giants detection}

Under the null hypothesis that the observed LV is drawn from the same population as the simulated volumes, the per-volume detection probability estimated from the simulations is $p=k/N$. With a single observed LV in which the CoG is detected ($N=1$, $k=1$), the one-sided $p$-value for observing at least one CoG equals $p_{\rm val}= p$, and the equivalent Gaussian significance  is
\begin{equation}
z=\Phi^{-1}(1-p_{\rm val}).
\end{equation}

where $\Phi$ is the Cumulative Distribution Function (CDF) of a normal distribution. Using the empirical frequencies measured in the three simulation sets, we obtain the $z$-score values shown in Table~\ref{tab:sigma}. The range of values takes into account the $\pm \sigma_{K\rightarrow200}$ region within the total scatter of the L+RP relation.

\begin{table}
\scriptsize
 \caption{Deviation in units of standard deviation ($\sigma$) between the LVG detection and $\Lambda$CDM predictions. Intervals denote the range of significance when considering model-dependent systematic errors.}
 \label{tab:sigma}
 \begin{center}
 \begin{tabular}{>{\centering\arraybackslash}m{1.5cm} >{\centering\arraybackslash}m{1.5cm} >{\centering\arraybackslash}m{1.5cm} >{\centering\arraybackslash}m{1.5cm} >{\centering\arraybackslash}m{1.5cm}}
 \hline\hline
  \noalign{\vspace {.1cm}}
 &HESTIA&SMD&SMD Random\\
 \hline
 \noalign{\vspace {.1cm}}
 Weak $\, \bar{n_{I}} = 0.15$&1.69-2.16 $\sigma$&2.05-2.73 $\sigma$&2.78-3.33 $\sigma$\\
  Moderate $\bar{n_{I}} = 0.20$&-&2.68-2.92 $\sigma$&3.22-3.70 $\sigma$\\
 
 \noalign{\vspace {.1cm}}
 \hline\hline
 \end{tabular}
 \end{center}
 \end{table}

Table~\ref{tab:sigma} reports the significance of the deviation between the observed CoG in LVG and the null hypothesis based on our simulated samples. The presence of a weak CoG-like ring in LVG corresponds to a $\gtrsim2.78\sigma$ anomaly relative to randomly chosen volumes, but this tension decreases to $\gtrsim2.05\sigma$ when comparing with volumes that contain an LG-like system, and falls to $\gtrsim1.69\sigma$ for volumes drawn from simulations with constrained initial conditions. In all cases, tension decreases as we reduce the effect of cosmic variance. 

In the case where CoG has a moderate statistical presence, similar to the detection in LVG, the tension is greater: adopting the SMD sample restricted to LG-like environments as a realistic analogue of the observed LV yields a deviation of $\gtrsim2.7\sigma$ from the $\rm \Lambda$CDM expectation.

\section{Discussion and conclusions}
\label{sec:conc}

To determine whether Council of Giants (CoG) formation is possible in the $\rm \Lambda$CDM model, we compared the observed Local Volume Galaxy survey (LVG) with its analogues extracted from two types of cosmological simulations: the Small MultiDark (SMD), based on random initial conditions and representative of the average cosmic environment, and the HESTIA simulations, which use constrained initial conditions to reproduce the large-scale structure of the Local Universe. In Paper~\hyperlink{part1}{I}, we introduced the HINORA algorithm to identify generic ring-like configurations in point distributions and confirmed the presence of the CoG in the LVG. Here, we extend that analysis to examine how frequently similar ring structures arise within these simulated reproductions of the Local Universe.

The simulated regions analysed in this work were selected to reproduce environments consistent with the observed Local Volume (LV). Using the known properties of the Local Group (LG) and the Virgo cluster as selection criteria, we identified 64 realisations from the HESTIA suite that successfully reproduce the local cosmography. Applying the same criteria to the SMD simulation yielded about four thousand comparable regions. The cumulative halo mass functions (CHMFs) of these samples show that both SMD and HESTIA reproduce the overall halo abundance observed in the LVG, although the observational data reveal a modest excess in the CoG mass regime ($M_{200} \sim 10^{11-12}M_{\odot}$). A similar trend is seen in the halo density contrast, with simulated LVs appearing slightly overdense relative to the cosmic background, consistent with the LG residing in an intermediate-density environment. These results confirm that our simulated volumes provide a realistic representation of the LV, yet they also suggest that the observed region is somewhat richer in massive halos than typical $\Lambda$CDM counterparts.

Using HINORA on the simulated samples, we recorded how many volumes contain a CoG-like ring with weak, moderate (i.e. comparable to the observation) or strong statistical presence, described by different values of the minimum halo fraction. Our main result is that CoG-like structures are very rare in the explored $\Lambda$CDM realisations, regardless of the detection threshold. Detection rates differ by more than an order of magnitude between the simulation sets, which provides useful information on the environmental mechanisms that can favour the appearance of such rings. In particular, when volumes are drawn completely at random from the SMD simulation box the probability of finding a CoG-like ring is very low (of order $\lesssim0.2\%$ depending on the presence threshold). Imposing LG-selection criteria in SMD increases the frequency to $\lesssim2\%$, while restricting the large-scale modes to match CosmicFlows-2 constraints (HESTIA) raises the weak-ring frequency further (up to $\sim 5\%$). However, the HESTIA constrained simulations do not reproduce a CoG with a statistical presence comparable to that observed in the LVG, although the appearance of CoG-like structures is slightly favoured in HESTIA compared to purely random realizations. With new observational data \citep{Tully2023_CF4} and improved algorithms for constructing constrained simulations \citep{Valade2026}, it will be possible to test again whether the observed velocity field predicts not only the Virgo cluster, the LG, and the Local Void, but also the CoG. Such a result would support the interpretation that the CoG is inherent to the specific environmental conditions of the LG.

Adopting the SMD sample restricted to LG-like environments as a conservative $\Lambda$CDM null hypothesis, the observed CoG in LVG departs from the model expectation at the level of $>2.7\sigma$. This level of tension indicates that, under the assumptions and detection criteria adopted in this work, the Local Universe is atypical compared to the ensemble of simulated MW-Andromeda regions, placing our galactic environment within the rare $\sim0.35\%$ of volumes that host a CoG-like ring.

Although this tension is mild, we can consider two possible solutions to this disagreement. The first is to attribute the existence of the CoG to chance. This perspective is fully valid, although future studies are still required to better determine the probability of CoG formation. In the present investigation we have focused on the geometrical properties that define the ring structure of the CoG, as well as on the mass of the galaxies that compose it. However, there are additional properties reported by \citet{McCall2014} that make the CoG special. Among them, the original paper points out that a large fraction of its galaxies are spirals, with only two being elliptical. Furthermore, based on the peculiar velocity dispersion, \citet{McCall2014} shows that the CoG appears to be a dynamically stable structure, as a typical member would require about 9.6 Gyr to leave the Local Sheet. Specifying these additional properties in the observations and considering them in (hydrodynamic) simulations could improve the probability estimation of CoG formation.

A second possible explanation for the tension, which should be interpreted with caution, is that the formation of the CoG involves physical processes not captured by our cosmological $\rm \Lambda$CDM simulations. Exploring hydrodynamic simulations could reveal whether this issue is related to baryonic matter interactions. The standard cosmological model, in the absence of hydrodynamical processes, does not include any known mechanisms capable of generating a $\sim$3.5~Mpc radius matter ring or the Local Sheet in which it resides \citep{Peebles2023}, so exploring models that might alleviate this tension is desirable. Alternative scenarios that can produce flat structures include topological defects that favor the emergence of Local Sheet-like configurations, such as cosmic strings \citep{Peebles2023} or the (a)symmetron \citep{Christiansen2024}. Investigating these and other possible extensions of $\rm \Lambda$CDM could help clarify the origin of the tension associated with the existence of the CoG.


\begin{acknowledgement}
All authors thank the referee for their constructive comments that helped to improve the paper. EO thanks Daniel Ceverino for his advice and Sabina Olex for her perseverance. AK likes to thank Sonic Youth for daydream nation. 
\end{acknowledgement}

\paragraph{Funding Statement}
EO and AK are supported by the Ministerio de Ciencia e Innovaci\'{o}n (MICINN) under research grant PID2021-122603NB-C21 as well as project PID2024-156100NB-C21 financed by MICIU /AEI/10.13039/5011000 11033/ FEDER, UE. EO received predoctoral fellowship from MICINN (FPI program, Ref. PRE2022-102254) The authors gratefully acknowledge the \href{https://urldefense.com/v3/__http://www.gauss-centre.eu__;!!D9dNQwwGXtA!UgLgyfHtHeLbJ7zfYIyi_a2eqJQgttCnrDMqo__k4F1YBkFOS0VdZuCexfZGlhnKBvwquow9D4XxdPZ4c-6OlAW66JdDmA$}{Gauss Centre for Supercomputing e.V} for funding this project by providing computing time on the GCS Supercomputer SuperMUC-NG at \href{https://urldefense.com/v3/__http://www.lrz.de__;!!D9dNQwwGXtA!UgLgyfHtHeLbJ7zfYIyi_a2eqJQgttCnrDMqo__k4F1YBkFOS0VdZuCexfZGlhnKBvwquow9D4XxdPZ4c-6OlAVMoHn2rw$}{Leibniz Supercomputing Centre}.


\paragraph{Data Availability Statement}

The \href{https://urldefense.com/v3/__http://www.cosmosim.org__;!!D9dNQwwGXtA!UgLgyfHtHeLbJ7zfYIyi_a2eqJQgttCnrDMqo__k4F1YBkFOS0VdZuCexfZGlhnKBvwquow9D4XxdPZ4c-6OlAVaS7wtOw$}{CosmoSim data base} provides access to the SMD simulation and the Rockstar halo data. The data base is a service by the Leibniz Institute for Astrophysics Potsdam (AIP). 

\printendnotes

\printbibliography

@ARTICLE{Libeskind2020,
       author = {{Libeskind}, Noam I. and {Carlesi}, Edoardo and {Grand}, Robert J.~J. and {Khalatyan}, Arman and {Knebe}, Alexander and {Pakmor}, Ruediger and {Pilipenko}, Sergey and {Pawlowski}, Marcel S. and {Sparre}, Martin and {Tempel}, Elmo and {Wang}, Peng and {Courtois}, H{\'e}l{\`e}ne M. and {Gottl{\"o}ber}, Stefan and {Hoffman}, Yehuda and {Minchev}, Ivan and {Pfrommer}, Christoph and {Sorce}, Jenny G. and {Springel}, Volker and {Steinmetz}, Matthias and {Tully}, R. Brent and {Vogelsberger}, Mark and {Yepes}, Gustavo},
        title = "{The HESTIA project: simulations of the Local Group}",
      journal = {mnras},
         year = 2020,
        month = oct,
       volume = {498},
       number = {2},
        pages = {2968-2983},
          doi = {10.1093/mnras/staa2541},
archivePrefix = {arXiv},
       eprint = {2008.04926},
 primaryClass = {astro-ph.GA},
       adsurl = {https://ui.adsabs.harvard.edu/abs/2020MNRAS.498.2968L}
}

@article{Bullock2017,
author = {Bullock, James S. and Boylan-Kolchin, Michael},
title = {Small-Scale Challenges to the $\Lambda$CDM Paradigm},
journal = {Annual Review of Astronomy and Astrophysics},
volume = {55},
number = {1},
pages = {343-387},
year = {2017},
doi = {10.1146/annurev-astro-091916-055313},

URL = { 
    
        https://doi.org/10.1146/annurev-astro-091916-055313
    
    

},
eprint = { 
    
        https://doi.org/10.1146/annurev-astro-091916-055313
    
    

}
}

@article{Tully2008,
doi = {10.1086/527428},
url = {https://dx.doi.org/10.1086/527428},
year = {2008},
month = {mar},
publisher = {},
volume = {676},
number = {1},
pages = {184},
author = {R. Brent Tully and Edward J. Shaya and Igor D. Karachentsev and Hélène M. Courtois and Dale D. Kocevski and Luca Rizzi and Alan Peel},
title = {Our Peculiar Motion Away from the Local Void},
journal = {The Astrophysical Journal}
}

@article{Karachentsev2013,
doi = {10.1088/0004-6256/145/4/101},
url = {https://dx.doi.org/10.1088/0004-6256/145/4/101},
year = {2013},
month = {mar},
publisher = {The American Astronomical Society},
volume = {145},
number = {4},
pages = {101},
author = {Igor D. Karachentsev and Dmitry I. Makarov and Elena I. Kaisina},
title = {UPDATED NEARBY GALAXY CATALOG},
journal = {The Astronomical Journal}
}

@article{Neuzil2020,
    author = {Neuzil, Maria K and Mansfield, Philip and Kravtsov, Andrey V},
    title = "{The Sheet of Giants: Unusual properties of the Milky Way’s immediate neighbourhood}",
    journal = {Monthly Notices of the Royal Astronomical Society},
    volume = {494},
    number = {2},
    pages = {2600-2617},
    year = {2020},
    month = {04},
    issn = {0035-8711},
    doi = {10.1093/mnras/staa898},
    url = {https://doi.org/10.1093/mnras/staa898},
    eprint = {https://academic.oup.com/mnras/article-pdf/494/2/2600/33113646/staa898.pdf},
}

@article{McCall2014,
    author = {McCall, Marshall L.},
    title = "{A Council of Giants}",
    journal = {Monthly Notices of the Royal Astronomical Society},
    volume = {440},
    number = {1},
    pages = {405-426},
    year = {2014},
    month = {02},
    issn = {0035-8711},
    doi = {10.1093/mnras/stu199},
    url = {https://doi.org/10.1093/mnras/stu199},
    eprint = {https://academic.oup.com/mnras/article-pdf/440/1/405/9376493/stu199.pdf},
}

@article{Bell2003,
doi = {10.1086/378847},
url = {https://dx.doi.org/10.1086/378847},
year = {2003},
month = {dec},
publisher = {},
volume = {149},
number = {2},
pages = {289},
author = {Eric F. Bell and Daniel H. McIntosh and Neal Katz and Martin D. Weinberg},
title = {The Optical and Near-Infrared Properties of Galaxies. I.
Luminosity and Stellar Mass Functions},
journal = {The Astrophysical Journal Supplement Series}
}

@article{Beare2019,
doi = {10.3847/1538-4357/ab041a},
url = {https://dx.doi.org/10.3847/1538-4357/ab041a},
year = {2019},
month = {mar},
publisher = {The American Astronomical Society},
volume = {873},
number = {1},
pages = {78},
author = {Richard Beare and Michael J. I. Brown and Kevin Pimbblet and Edward N. Taylor},
title = {Evolution of the Stellar Mass Function and Infrared Luminosity Function of Galaxies since z=1.2},
journal = {The Astrophysical Journal}
}

@ARTICLE{Springel2005,
       author = {{Springel}, Volker and {White}, Simon D.~M. and {Jenkins}, Adrian and {Frenk}, Carlos S. and {Yoshida}, Naoki and {Gao}, Liang and {Navarro}, Julio and {Thacker}, Robert and {Croton}, Darren and {Helly}, John and {Peacock}, John A. and {Cole}, Shaun and {Thomas}, Peter and {Couchman}, Hugh and {Evrard}, August and {Colberg}, J{\"o}rg and {Pearce}, Frazer},
        title = "{Simulations of the formation, evolution and clustering of galaxies and quasars}",
      journal = {Nature},
         year = 2005,
        month = jun,
       volume = {435},
       number = {7042},
        pages = {629-636},
          doi = {10.1038/nature03597},
archivePrefix = {arXiv},
       eprint = {astro-ph/0504097},
 primaryClass = {astro-ph},
       adsurl = {https://ui.adsabs.harvard.edu/abs/2005Natur.435..629S}
}

@ARTICLE{Hoffman1991,
       author = {{Hoffman}, Yehuda and {Ribak}, Erez},
        title = "{Constrained Realizations of Gaussian Fields: A Simple Algorithm}",
      journal = {The Astrophysical Journal Letters},
         year = 1991,
        month = oct,
       volume = {380},
        pages = {L5},
          doi = {10.1086/186160},
       adsurl = {https://ui.adsabs.harvard.edu/abs/1991ApJ...380L...5H}
}

@article{Tully2023_CF4,
doi = {10.3847/1538-4357/ac94d8},
url = {https://dx.doi.org/10.3847/1538-4357/ac94d8},
year = {2023},
month = {feb},
publisher = {The American Astronomical Society},
volume = {944},
number = {1},
pages = {94},
author = {R. Brent Tully and Ehsan Kourkchi and Hélène M. Courtois and Gagandeep S. Anand and John P. Blakeslee and Dillon Brout and Thomas de Jaeger and Alexandra Dupuy and Daniel Guinet and Cullan Howlett and Joseph B. Jensen and Daniel Pomarède and Luca Rizzi and David Rubin and Khaled Said and Daniel Scolnic and Benjamin E. Stahl},
title = {Cosmicflows-4},
journal = {The Astrophysical Journal}
}

@article{Puebla2017,
    author = {Rodríguez-Puebla, Aldo and Primack, Joel R. and Avila-Reese, Vladimir and Faber, S. M.},
    title = "{Constraining the galaxy–halo connection over the last 13.3 Gyr: star formation histories, galaxy mergers and structural properties}",
    journal = {Monthly Notices of the Royal Astronomical Society},
    volume = {470},
    number = {1},
    pages = {651-687},
    year = {2017},
    month = {05},
    issn = {0035-8711},
    doi = {10.1093/mnras/stx1172},
    url = {https://doi.org/10.1093/mnras/stx1172},
    eprint = {https://academic.oup.com/mnras/article-pdf/470/1/651/17845442/stx1172.pdf},
}

@ARTICLE{Olex2024,
       author = {{Olex}, Edward and \hypertarget{part1}{Knebe}, Alexander and {Libeskind}, Noam I. and {Makarov}, Dmitry I. and {Gottl{\"o}ber}, Stefan},
        title = "{HINORA, a method for detecting ring-like structures in 3D point distributions I: Application to the Local Volume Galaxy catalogue}",
      journal = {PASA},
         year = 2024,
        month = jan,
       volume = {41},
          eid = {e018},
        pages = {e018},
          doi = {10.1017/pasa.2024.21},
archivePrefix = {arXiv},
       eprint = {2403.06187},
 primaryClass = {astro-ph.CO},
       adsurl = {https://ui.adsabs.harvard.edu/abs/2024PASA...41...18O}
}

@ARTICLE{Tully2013,
       author = {{Tully}, R. Brent and {Courtois}, H{\'e}l{\`e}ne M. and {Dolphin}, Andrew E. and {Fisher}, J. Richard and {H{\'e}raudeau}, Philippe and {Jacobs}, Bradley A. and {Karachentsev}, Igor D. and {Makarov}, Dmitry and {Makarova}, Lidia and {Mitronova}, Sofia and {Rizzi}, Luca and {Shaya}, Edward J. and {Sorce}, Jenny G. and {Wu}, Po-Feng},
        title = "{Cosmicflows-2: The Data}",
      journal = {AJ},
         year = 2013,
        month = oct,
       volume = {146},
       number = {4},
          eid = {86},
        pages = {86},
          doi = {10.1088/0004-6256/146/4/86},
archivePrefix = {arXiv},
       eprint = {1307.7213},
 primaryClass = {astro-ph.CO},
       adsurl = {https://ui.adsabs.harvard.edu/abs/2013AJ....146...86T}
}

@ARTICLE{Klypin2016,
       author = {{Klypin}, Anatoly and {Yepes}, Gustavo and {Gottl{\"o}ber}, Stefan and {Prada}, Francisco and {He{\ss}}, Steffen},
        title = "{MultiDark simulations: the story of dark matter halo concentrations and density profiles}",
      journal = {MNRAS},
         year = 2016,
        month = apr,
       volume = {457},
       number = {4},
        pages = {4340-4359},
          doi = {10.1093/mnras/stw248},
archivePrefix = {arXiv},
       eprint = {1411.4001},
 primaryClass = {astro-ph.CO},
       adsurl = {https://ui.adsabs.harvard.edu/abs/2016MNRAS.457.4340K}
}

@ARTICLE{Knollmann2009,
       author = {{Knollmann}, Steffen R. and {Knebe}, Alexander},
        title = "{AHF: Amiga's Halo Finder}",
      journal = {APJs},
         year = 2009,
        month = jun,
       volume = {182},
       number = {2},
        pages = {608-624},
          doi = {10.1088/0067-0049/182/2/608},
archivePrefix = {arXiv},
       eprint = {0904.3662},
 primaryClass = {astro-ph.CO},
       adsurl = {https://ui.adsabs.harvard.edu/abs/2009ApJS..182..608K}
}

@ARTICLE{Weinberger2020,
       author = {{Weinberger}, Rainer and {Springel}, Volker and {Pakmor}, R{\"u}diger},
        title = "{The AREPO Public Code Release}",
      journal = {APJs},
         year = 2020,
        month = jun,
       volume = {248},
       number = {2},
          eid = {32},
        pages = {32},
          doi = {10.3847/1538-4365/ab908c},
archivePrefix = {arXiv},
       eprint = {1909.04667},
 primaryClass = {astro-ph.IM},
       adsurl = {https://ui.adsabs.harvard.edu/abs/2020ApJS..248...32W}
}

@ARTICLE{Into2013,
       author = {{Into}, Tom and {Portinari}, Laura},
        title = "{New colour-mass-to-light relations: the role of the asymptotic giant branch phase and of interstellar dust}",
      journal = {MNRAS},
         year = 2013,
        month = apr,
       volume = {430},
       number = {4},
        pages = {2715-2731},
          doi = {10.1093/mnras/stt071},
archivePrefix = {arXiv},
       eprint = {1301.2178},
 primaryClass = {astro-ph.CO},
       adsurl = {https://ui.adsabs.harvard.edu/abs/2013MNRAS.430.2715I}
}

@ARTICLE{Kourkchi2017,
       author = {{Kourkchi}, Ehsan and {Tully}, R. Brent},
        title = "{Galaxy Groups Within 3500 km s$^{-1}$}",
      journal = {APJ},
         year = 2017,
        month = jul,
       volume = {843},
       number = {1},
          eid = {16},
        pages = {16},
          doi = {10.3847/1538-4357/aa76db},
archivePrefix = {arXiv},
       eprint = {1705.08068},
 primaryClass = {astro-ph.GA},
       adsurl = {https://ui.adsabs.harvard.edu/abs/2017ApJ...843...16K}
}

@ARTICLE{Karachentsev2021,
       author = {{Karachentsev}, Igor and {Kashibadze}, Olga},
        title = "{Tracing the local volume galaxy halo-to-stellar mass ratio with satellite kinematics}",
      journal = {Astronomische Nachrichten},
         year = 2021,
        month = aug,
       volume = {342},
       number = {999},
        pages = {999-1023},
          doi = {10.1002/asna.20210018},
archivePrefix = {arXiv},
       eprint = {2109.00336},
 primaryClass = {astro-ph.GA},
       adsurl = {https://ui.adsabs.harvard.edu/abs/2021AN....342..999K}
}

@ARTICLE{White2001,
       author = {{White}, M.},
        title = "{The mass of a halo}",
      journal = {A\&A},
         year = 2001,
        month = feb,
       volume = {367},
        pages = {27-32},
          doi = {10.1051/0004-6361:20000357},
archivePrefix = {arXiv},
       eprint = {astro-ph/0011495},
 primaryClass = {astro-ph},
       adsurl = {https://ui.adsabs.harvard.edu/abs/2001A&A...367...27W}
}

@ARTICLE{Pawlowski2018,
       author = {{Pawlowski}, Marcel S.},
        title = "{The planes of satellite galaxies problem, suggested solutions, and open questions}",
      journal = {Modern Physics Letters A},
         year = 2018,
        month = feb,
       volume = {33},
       number = {6},
          eid = {1830004},
        pages = {1830004},
          doi = {10.1142/S0217732318300045},
archivePrefix = {arXiv},
       eprint = {1802.02579},
 primaryClass = {astro-ph.GA},
       adsurl = {https://ui.adsabs.harvard.edu/abs/2018MPLA...3330004P}
}

@ARTICLE{Matias2024,
       author = {{G{\'a}mez-Mar{\'\i}n}, Mat{\'\i}as and {Santos-Santos}, Isabel and {Dom{\'\i}nguez-Tenreiro}, Rosa and {Pedrosa}, Susana E. and {Tissera}, Patricia B. and {G{\'o}mez-Flechoso}, M. {\'A}ngeles and {Artal}, H{\'e}ctor},
        title = "{The Origin of Kinematically Persistent Planes of Satellites as Driven by the Early Evolution of the Cosmic Web in {\ensuremath{\Lambda}}CDM}",
      journal = {APJ},
         year = 2024,
        month = apr,
       volume = {965},
       number = {2},
          eid = {154},
        pages = {154},
          doi = {10.3847/1538-4357/ad27da},
archivePrefix = {arXiv},
       eprint = {2402.03288},
 primaryClass = {astro-ph.GA},
       adsurl = {https://ui.adsabs.harvard.edu/abs/2024ApJ...965..154G}
}

@ARTICLE{Libeskind2009,
       author = {{Libeskind}, Noam I. and {Frenk}, Carlos S. and {Cole}, Shaun and {Jenkins}, Adrian and {Helly}, John C.},
        title = "{How common is the Milky Way-satellite system alignment?}",
      journal = {MNRAS},
         year = 2009,
        month = oct,
       volume = {399},
       number = {2},
        pages = {550-558},
          doi = {10.1111/j.1365-2966.2009.15315.x},
archivePrefix = {arXiv},
       eprint = {0905.1696},
 primaryClass = {astro-ph.CO},
       adsurl = {https://ui.adsabs.harvard.edu/abs/2009MNRAS.399..550L}
}

@ARTICLE{Libeskind2005,
       author = {{Libeskind}, Noam I. and {Frenk}, Carlos S. and {Cole}, Shaun and {Helly}, John C. and {Jenkins}, Adrian and {Navarro}, Julio F. and {Power}, Chris},
        title = "{The distribution of satellite galaxies: the great pancake}",
      journal = {MNRAS},
         year = 2005,
        month = oct,
       volume = {363},
       number = {1},
        pages = {146-152},
          doi = {10.1111/j.1365-2966.2005.09425.x},
archivePrefix = {arXiv},
       eprint = {astro-ph/0503400},
 primaryClass = {astro-ph},
       adsurl = {https://ui.adsabs.harvard.edu/abs/2005MNRAS.363..146L}
}

@ARTICLE{Lovell2011,
       author = {{Lovell}, Mark R. and {Eke}, Vincent R. and {Frenk}, Carlos S. and {Jenkins}, Adrian},
        title = "{The link between galactic satellite orbits and subhalo accretion}",
      journal = {MNRAS},
         year = 2011,
        month = jun,
       volume = {413},
       number = {4},
        pages = {3013-3021},
          doi = {10.1111/j.1365-2966.2011.18377.x},
archivePrefix = {arXiv},
       eprint = {1008.0484},
 primaryClass = {astro-ph.CO},
       adsurl = {https://ui.adsabs.harvard.edu/abs/2011MNRAS.413.3013L}
}

@ARTICLE{Wang2013,
       author = {{Wang}, Jie and {Frenk}, Carlos S. and {Cooper}, Andrew P.},
        title = "{The spatial distribution of galactic satellites in the {\ensuremath{\Lambda}} cold dark matter cosmology}",
      journal = {MNRAS},
         year = 2013,
        month = feb,
       volume = {429},
       number = {2},
        pages = {1502-1513},
          doi = {10.1093/mnras/sts442},
archivePrefix = {arXiv},
       eprint = {1206.1340},
 primaryClass = {astro-ph.GA},
       adsurl = {https://ui.adsabs.harvard.edu/abs/2013MNRAS.429.1502W}
}

@ARTICLE{Cautun2015,
       author = {{Cautun}, Marius and {Bose}, Sownak and {Frenk}, Carlos S. and {Guo}, Qi and {Han}, Jiaxin and {Hellwing}, Wojciech A. and {Sawala}, Till and {Wang}, Wenting},
        title = "{Planes of satellite galaxies: when exceptions are the rule}",
      journal = {MNRAS},
         year = 2015,
        month = oct,
       volume = {452},
       number = {4},
        pages = {3838-3852},
          doi = {10.1093/mnras/stv1557},
archivePrefix = {arXiv},
       eprint = {1506.04151},
 primaryClass = {astro-ph.GA},
       adsurl = {https://ui.adsabs.harvard.edu/abs/2015MNRAS.452.3838C}
}

@ARTICLE{Buck2015,
       author = {{Buck}, Tobias and {Macci{\`o}}, Andrea V. and {Dutton}, Aaron A.},
        title = "{Evidence for Early Filamentary Accretion from the Andromeda Galaxy{\textquoteright}s Thin Plane of Satellites}",
      journal = {APJ},
         year = 2015,
        month = aug,
       volume = {809},
       number = {1},
          eid = {49},
        pages = {49},
          doi = {10.1088/0004-637X/809/1/49},
archivePrefix = {arXiv},
       eprint = {1504.05193},
 primaryClass = {astro-ph.GA},
       adsurl = {https://ui.adsabs.harvard.edu/abs/2015ApJ...809...49B}
}

@ARTICLE{Gillet2015,
       author = {{Gillet}, N. and {Ocvirk}, P. and {Aubert}, D. and {Knebe}, A. and {Libeskind}, N. and {Yepes}, G. and {Gottl{\"o}ber}, S. and {Hoffman}, Y.},
        title = "{Vast Planes of Satellites in a High-resolution Simulation of the Local Group: Comparison to Andromeda}",
      journal = {APJ},
         year = 2015,
        month = feb,
       volume = {800},
       number = {1},
          eid = {34},
        pages = {34},
          doi = {10.1088/0004-637X/800/1/34},
archivePrefix = {arXiv},
       eprint = {1412.3110},
 primaryClass = {astro-ph.GA},
       adsurl = {https://ui.adsabs.harvard.edu/abs/2015ApJ...800...34G}
}

@ARTICLE{Ahmed2017,
       author = {{Ahmed}, Sheehan H. and {Brooks}, Alyson M. and {Christensen}, Charlotte R.},
        title = "{The role of baryons in creating statistically significant planes of satellites around Milky Way-mass galaxies}",
      journal = {MNRAS},
         year = 2017,
        month = apr,
       volume = {466},
       number = {3},
        pages = {3119-3132},
          doi = {10.1093/mnras/stw3271},
archivePrefix = {arXiv},
       eprint = {1610.03077},
 primaryClass = {astro-ph.GA},
       adsurl = {https://ui.adsabs.harvard.edu/abs/2017MNRAS.466.3119A}
}

@ARTICLE{Shao2019,
       author = {{Shao}, Shi and {Cautun}, Marius and {Frenk}, Carlos S.},
        title = "{Evolution of galactic planes of satellites in the EAGLE simulation}",
      journal = {MNRAS},
         year = 2019,
        month = sep,
       volume = {488},
       number = {1},
        pages = {1166-1179},
          doi = {10.1093/mnras/stz1741},
archivePrefix = {arXiv},
       eprint = {1904.02719},
 primaryClass = {astro-ph.GA},
       adsurl = {https://ui.adsabs.harvard.edu/abs/2019MNRAS.488.1166S}
}

@ARTICLE{Samuel2021,
       author = {{Samuel}, Jenna and {Wetzel}, Andrew and {Chapman}, Sierra and {Tollerud}, Erik and {Hopkins}, Philip F. and {Boylan-Kolchin}, Michael and {Bailin}, Jeremy and {Faucher-Gigu{\`e}re}, Claude-Andr{\'e}},
        title = "{Planes of satellites around Milky Way/M31-mass galaxies in the FIRE simulations and comparisons with the Local Group}",
      journal = {MNRAS},
         year = 2021,
        month = jun,
       volume = {504},
       number = {1},
        pages = {1379-1397},
          doi = {10.1093/mnras/stab955},
archivePrefix = {arXiv},
       eprint = {2010.08571},
 primaryClass = {astro-ph.GA},
       adsurl = {https://ui.adsabs.harvard.edu/abs/2021MNRAS.504.1379S}
}

@ARTICLE{Pham2023,
       author = {{Pham}, Khanh and {Kravtsov}, Andrey and {Manwadkar}, Viraj},
        title = "{Spatial and orbital planes of the Milky Way satellites: unusual but consistent with {\ensuremath{\Lambda}}CDM}",
      journal = {MNRAS},
         year = 2023,
        month = apr,
       volume = {520},
       number = {3},
        pages = {3937-3946},
          doi = {10.1093/mnras/stad335},
archivePrefix = {arXiv},
       eprint = {2209.02714},
 primaryClass = {astro-ph.GA},
       adsurl = {https://ui.adsabs.harvard.edu/abs/2023MNRAS.520.3937P}
}

@ARTICLE{Rubin1951,
       author = {{Rubin}, Vera Cooper},
        title = "{Differential rotation of the inner metagalaxy.}",
      journal = {AJ},
         year = 1951,
        month = jan,
       volume = {56},
        pages = {47},
          doi = {10.1086/106628},
       adsurl = {https://ui.adsabs.harvard.edu/abs/1951AJ.....56S..47R}
}

@ARTICLE{Vaucouleurs1953,
       author = {{de Vaucouleurs}, Gerard},
        title = "{Evidence for a local super,galaxy}",
      journal = {AJ},
         year = 1953,
        month = feb,
       volume = {58},
        pages = {30},
          doi = {10.1086/106805},
       adsurl = {https://ui.adsabs.harvard.edu/abs/1953AJ.....58...30D}
}

@ARTICLE{Vaucouleurs1958,
       author = {{de Vaucouleurs}, Gerard},
        title = "{Further evidence for a local super-cluster of galaxies: rotation and expansion}",
      journal = {AJ},
         year = 1958,
        month = jul,
       volume = {63},
        pages = {253},
          doi = {10.1086/107742},
       adsurl = {https://ui.adsabs.harvard.edu/abs/1958AJ.....63..253D}
}

@ARTICLE{Peebles2023,
       author = {{Peebles}, P.~J.~E.},
        title = "{Flat patterns in cosmic structure}",
      journal = {MNRAS},
         year = 2023,
        month = dec,
       volume = {526},
       number = {3},
        pages = {4490-4501},
          doi = {10.1093/mnras/stad3051},
archivePrefix = {arXiv},
       eprint = {2308.04245},
 primaryClass = {astro-ph.CO},
       adsurl = {https://ui.adsabs.harvard.edu/abs/2023MNRAS.526.4490P}
}

@ARTICLE{Calvo2023,
       author = {{Aragon-Calvo}, M.~A. and {Silk}, Joseph and {Neyrinck}, Mark},
        title = "{The unusual Milky Way-local sheet system: implications for spin strength and alignment}",
      journal = {MNRAS},
         year = 2023,
        month = mar,
       volume = {520},
       number = {1},
        pages = {L28-L32},
          doi = {10.1093/mnrasl/slac161},
archivePrefix = {arXiv},
       eprint = {2208.03338},
 primaryClass = {astro-ph.GA},
       adsurl = {https://ui.adsabs.harvard.edu/abs/2023MNRAS.520L..28A}
}

@ARTICLE{Puebla2015,
       author = {{Rodr{\'\i}guez-Puebla}, Aldo and {Avila-Reese}, Vladimir and {Yang}, Xiaohu and {Foucaud}, Sebastien and {Drory}, Niv and {Jing}, Y.~P.},
        title = "{The Stellar-to-Halo Mass Relation of Local Galaxies Segregates by Color}",
      journal = {APJ},
         year = 2015,
        month = feb,
       volume = {799},
       number = {2},
          eid = {130},
        pages = {130},
          doi = {10.1088/0004-637X/799/2/130},
archivePrefix = {arXiv},
       eprint = {1408.5407},
 primaryClass = {astro-ph.GA},
       adsurl = {https://ui.adsabs.harvard.edu/abs/2015ApJ...799..130R}
}

@ARTICLE{Christiansen2024,
       author = {{Christiansen}, {\O}yvind and {Hassani}, Farbod and {Mota}, David F.},
        title = "{Asimulation: Domain formation and impact on observables in resolved cosmological simulations of the (a)symmetron}",
      journal = {A\&A},
         year = 2024,
        month = sep,
       volume = {689},
          eid = {A6},
        pages = {A6},
          doi = {10.1051/0004-6361/202449188},
archivePrefix = {arXiv},
       eprint = {2401.02410},
 primaryClass = {astro-ph.CO},
       adsurl = {https://ui.adsabs.harvard.edu/abs/2024A&A...689A...6C}
}

@ARTICLE{Rock2015,
       author = {{R{\"o}ck}, B. and {Vazdekis}, A. and {Peletier}, R.~F. and {Knapen}, J.~H. and {Falc{\'o}n-Barroso}, J.},
        title = "{Stellar population synthesis models between 2.5 and 5 {\ensuremath{\mu}}m based on the empirical IRTF stellar library}",
      journal = {MNRAS},
         year = 2015,
        month = may,
       volume = {449},
       number = {3},
        pages = {2853-2874},
          doi = {10.1093/mnras/stv503},
archivePrefix = {arXiv},
       eprint = {1505.01837},
 primaryClass = {astro-ph.GA},
       adsurl = {https://ui.adsabs.harvard.edu/abs/2015MNRAS.449.2853R}
}

@ARTICLE{Wen2013,
       author = {{Wen}, Xiao-Qing and {Wu}, Hong and {Zhu}, Yi-Nan and {Lam}, Man I. and {Wu}, Chao-Jian and {Wicker}, James and {Zhao}, Yong-Heng},
        title = "{The stellar masses of galaxies from the 3.4 {\ensuremath{\mu}}m band of the WISE All-Sky Survey}",
      journal = {MNRAS},
         year = 2013,
        month = aug,
       volume = {433},
       number = {4},
        pages = {2946-2957},
          doi = {10.1093/mnras/stt939},
       adsurl = {https://ui.adsabs.harvard.edu/abs/2013MNRAS.433.2946W}
}

@ARTICLE{Meidt2014,
       author = {{Meidt}, Sharon E. and {Schinnerer}, Eva and {van de Ven}, Glenn and {Zaritsky}, Dennis and {Peletier}, Reynier and {Knapen}, Johan H. and {Sheth}, Kartik and {Regan}, Michael and {Querejeta}, Miguel and {Mu{\~n}oz-Mateos}, Juan-Carlos and {Kim}, Taehyun and {Hinz}, Joannah L. and {Gil de Paz}, Armando and {Athanassoula}, E. and {Bosma}, Albert and {Buta}, Ronald J. and {Cisternas}, Mauricio and {Ho}, Luis C. and {Holwerda}, Benne and {Skibba}, Ramin and {Laurikainen}, E. and {Salo}, H. and {Gadotti}, D.~A. and {Laine}, Jarkko and {Erroz-Ferrer}, S. and {Comer{\'o}n}, S{\'e}bastien and {Men{\'e}ndez-Delmestre}, K. and {Seibert}, M. and {Mizusawa}, T.},
        title = "{Reconstructing the Stellar Mass Distributions of Galaxies Using S$^{4}$G IRAC 3.6 and 4.5 {\ensuremath{\mu}}m Images. II. The Conversion from Light to Mass}",
      journal = {APJ},
         year = 2014,
        month = jun,
       volume = {788},
       number = {2},
          eid = {144},
        pages = {144},
          doi = {10.1088/0004-637X/788/2/144},
archivePrefix = {arXiv},
       eprint = {1402.5210},
 primaryClass = {astro-ph.GA},
       adsurl = {https://ui.adsabs.harvard.edu/abs/2014ApJ...788..144M}
}

@ARTICLE{Brinchmann2000,
       author = {{Brinchmann}, Jarle and {Ellis}, Richard S.},
        title = "{The Mass Assembly and Star Formation Characteristics of Field Galaxies of Known Morphology}",
      journal = {APJL},
         year = 2000,
        month = jun,
       volume = {536},
       number = {2},
        pages = {L77-L80},
          doi = {10.1086/312738},
archivePrefix = {arXiv},
       eprint = {astro-ph/0005120},
 primaryClass = {astro-ph},
       adsurl = {https://ui.adsabs.harvard.edu/abs/2000ApJ...536L..77B}
}

@ARTICLE{Cole2001,
       author = {{Cole}, Shaun and {Norberg}, Peder and {Baugh}, Carlton M. and {Frenk}, Carlos S. and {Bland-Hawthorn}, Joss and {Bridges}, Terry and {Cannon}, Russell and {Colless}, Matthew and {Collins}, Chris and {Couch}, Warrick and {Cross}, Nicholas and {Dalton}, Gavin and {De Propris}, Roberto and {Driver}, Simon P. and {Efstathiou}, George and {Ellis}, Richard S. and {Glazebrook}, Karl and {Jackson}, Carole and {Lahav}, Ofer and {Lewis}, Ian and {Lumsden}, Stuart and {Maddox}, Steve and {Madgwick}, Darren and {Peacock}, John A. and {Peterson}, Bruce A. and {Sutherland}, Will and {Taylor}, Keith},
        title = "{The 2dF galaxy redshift survey: near-infrared galaxy luminosity functions}",
      journal = {MNRAS},
         year = 2001,
        month = sep,
       volume = {326},
       number = {1},
        pages = {255-273},
          doi = {10.1046/j.1365-8711.2001.04591.x},
archivePrefix = {arXiv},
       eprint = {astro-ph/0012429},
 primaryClass = {astro-ph},
       adsurl = {https://ui.adsabs.harvard.edu/abs/2001MNRAS.326..255C}
}

@ARTICLE{Bundy2005,
       author = {{Bundy}, Kevin and {Ellis}, Richard S. and {Conselice}, Christopher J.},
        title = "{The Mass Assembly Histories of Galaxies of Various Morphologies in the GOODS Fields}",
      journal = {APJ},
         year = 2005,
        month = jun,
       volume = {625},
       number = {2},
        pages = {621-632},
          doi = {10.1086/429549},
archivePrefix = {arXiv},
       eprint = {astro-ph/0502204},
 primaryClass = {astro-ph},
       adsurl = {https://ui.adsabs.harvard.edu/abs/2005ApJ...625..621B}
}

@ARTICLE{Taylor2011,
       author = {{Taylor}, Edward N. and {Hopkins}, Andrew M. and {Baldry}, Ivan K. and {Brown}, Michael J.~I. and {Driver}, Simon P. and {Kelvin}, Lee S. and {Hill}, David T. and {Robotham}, Aaron S.~G. and {Bland-Hawthorn}, Joss and {Jones}, D.~H. and {Sharp}, R.~G. and {Thomas}, Daniel and {Liske}, Jochen and {Loveday}, Jon and {Norberg}, Peder and {Peacock}, J.~A. and {Bamford}, Steven P. and {Brough}, Sarah and {Colless}, Matthew and {Cameron}, Ewan and {Conselice}, Christopher J. and {Croom}, Scott M. and {Frenk}, C.~S. and {Gunawardhana}, Madusha and {Kuijken}, Konrad and {Nichol}, R.~C. and {Parkinson}, H.~R. and {Phillipps}, S. and {Pimbblet}, K.~A. and {Popescu}, C.~C. and {Prescott}, Matthew and {Sutherland}, W.~J. and {Tuffs}, R.~J. and {van Kampen}, Eelco and {Wijesinghe}, D.},
        title = "{Galaxy And Mass Assembly (GAMA): stellar mass estimates}",
      journal = {MNRAS},
         year = 2011,
        month = dec,
       volume = {418},
       number = {3},
        pages = {1587-1620},
          doi = {10.1111/j.1365-2966.2011.19536.x},
archivePrefix = {arXiv},
       eprint = {1108.0635},
 primaryClass = {astro-ph.CO},
       adsurl = {https://ui.adsabs.harvard.edu/abs/2011MNRAS.418.1587T}
}

@ARTICLE{McGaugh2014,
       author = {{McGaugh}, Stacy S. and {Schombert}, James M.},
        title = "{Color-Mass-to-light-ratio Relations for Disk Galaxies}",
      journal = {AJ},
         year = 2014,
        month = nov,
       volume = {148},
       number = {5},
          eid = {77},
        pages = {77},
          doi = {10.1088/0004-6256/148/5/77},
archivePrefix = {arXiv},
       eprint = {1407.1839},
 primaryClass = {astro-ph.GA},
       adsurl = {https://ui.adsabs.harvard.edu/abs/2014AJ....148...77M}
}

@ARTICLE{Lelli2016,
       author = {{Lelli}, Federico and {McGaugh}, Stacy S. and {Schombert}, James M.},
        title = "{SPARC: Mass Models for 175 Disk Galaxies with Spitzer Photometry and Accurate Rotation Curves}",
      journal = {AJ},
         year = 2016,
        month = dec,
       volume = {152},
       number = {6},
          eid = {157},
        pages = {157},
          doi = {10.3847/0004-6256/152/6/157},
archivePrefix = {arXiv},
       eprint = {1606.09251},
 primaryClass = {astro-ph.GA},
       adsurl = {https://ui.adsabs.harvard.edu/abs/2016AJ....152..157L}
}

@ARTICLE{Kim2025,
       author = {{Kim}, Taehyun and {Kim}, Minjin and {Ho}, Luis C. and {Li}, Yang A. and {Jeong}, Woong-Seob and {Kim}, Dohyeong and {Kim}, Yongjung and {Lee}, Bomee and {Lee}, Dongseob and {Lee}, Jeong Hwan and {Pyo}, Jeonghyun and {Shim}, Hyunjin and {Son}, Suyeon and {Song}, Hyunmi and {Yang}, Yujin},
        title = "{Accuracy of Stellar Mass-to-light Ratios of Nearby Galaxies in the Near Infrared}",
      journal = {AJ},
         year = 2025,
        month = jan,
       volume = {169},
       number = {1},
          eid = {44},
        pages = {44},
          doi = {10.3847/1538-3881/ad95eb},
archivePrefix = {arXiv},
       eprint = {2411.10981},
 primaryClass = {astro-ph.GA},
       adsurl = {https://ui.adsabs.harvard.edu/abs/2025AJ....169...44K}
}

@ARTICLE{Wempe2025,
       author = {{Wempe}, Ewoud and {Helmi}, Amina and {White}, Simon D.~M. and {Jasche}, Jens and {Lavaux}, Guilhem},
        title = "{The effect of environment on the mass assembly history of the Milky Way and M31}",
      journal = {arXiv e-prints},
         year = 2025,
        month = jan,
          eid = {arXiv:2501.08089},
        pages = {arXiv:2501.08089},
          doi = {10.48550/arXiv.2501.08089},
archivePrefix = {arXiv},
       eprint = {2501.08089},
 primaryClass = {astro-ph.GA},
       adsurl = {https://ui.adsabs.harvard.edu/abs/2025arXiv250108089W}
}

@ARTICLE{ForeroRomero2015,
       author = {{Forero-Romero}, J.~E. and {Gonz{\'a}lez}, R.},
        title = "{The Local Group in the Cosmic Web}",
      journal = {\apj},
         year = 2015,
        month = jan,
       volume = {799},
       number = {1},
          eid = {45},
        pages = {45},
          doi = {10.1088/0004-637X/799/1/45},
archivePrefix = {arXiv},
       eprint = {1408.3166},
 primaryClass = {astro-ph.CO},
       adsurl = {https://ui.adsabs.harvard.edu/abs/2015ApJ...799...45F}
}

@ARTICLE{Katz1993,
       author = {{Katz}, Neal and {White}, Simon D.~M.},
        title = "{Hierarchical Galaxy Formation: Overmerging and the Formation of an X-Ray Cluster}",
      journal = {\apj},
         year = 1993,
        month = aug,
       volume = {412},
        pages = {455},
          doi = {10.1086/172935},
       adsurl = {https://ui.adsabs.harvard.edu/abs/1993ApJ...412..455K}
}

@ARTICLE{Zaroubi1995,
       author = {{Zaroubi}, S. and {Hoffman}, Y. and {Fisher}, K.~B. and {Lahav}, O.},
        title = "{Wiener Reconstruction of the Large-Scale Structure}",
      journal = {\apj},
         year = 1995,
        month = aug,
       volume = {449},
        pages = {446},
          doi = {10.1086/176070},
archivePrefix = {arXiv},
       eprint = {astro-ph/9410080},
 primaryClass = {astro-ph},
       adsurl = {https://ui.adsabs.harvard.edu/abs/1995ApJ...449..446Z}
}

@INCOLLECTION{Hoffman2009,
       author = {{Hoffman}, Y.},
        title = "{Gaussian Fields and Constrained Simulations of the Large-Scale Structure}",
    booktitle = {Data Analysis in Cosmology},
         year = 2009,
       editor = {{Mart{\'\i}nez}, V.~J. and {Saar}, E. and {Mart{\'\i}nez-Gonz{\'a}lez}, E. and {Pons-Border{\'\i}a}, M. -J.},
       volume = {665},
        pages = {565-583},
          doi = {10.1007/978-3-540-44767-2_17},
       adsurl = {https://ui.adsabs.harvard.edu/abs/2009LNP...665..565H}
}

@ARTICLE{Gottloeber2010,
       author = {{Gottl\"ober}, Stefan and {Hoffman}, Yehuda and {Yepes}, Gustavo},
        title = "{Constrained Local UniversE Simulations (CLUES)}",
      journal = {arXiv e-prints},
         year = 2010,
        month = may,
          eid = {arXiv:1005.2687},
        pages = {arXiv:1005.2687},
          doi = {10.48550/arXiv.1005.2687},
archivePrefix = {arXiv},
       eprint = {1005.2687},
 primaryClass = {astro-ph.CO},
       adsurl = {https://ui.adsabs.harvard.edu/abs/2010arXiv1005.2687G}
}

@ARTICLE{Libeskind2010,
       author = {{Libeskind}, Noam I. and {Yepes}, Gustavo and {Knebe}, Alexander and {Gottl{\"o}ber}, Stefan and {Hoffman}, Yehuda and {Knollmann}, Steffen R.},
        title = "{Constrained simulations of the Local Group: on the radial distribution of substructures}",
      journal = {\mnras},
         year = 2010,
        month = jan,
       volume = {401},
       number = {3},
        pages = {1889-1897},
          doi = {10.1111/j.1365-2966.2009.15766.x},
archivePrefix = {arXiv},
       eprint = {0909.4423},
 primaryClass = {astro-ph.CO},
       adsurl = {https://ui.adsabs.harvard.edu/abs/2010MNRAS.401.1889L}
}

@ARTICLE{Carlesi2016,
       author = {{Carlesi}, Edoardo and {Sorce}, Jenny G. and {Hoffman}, Yehuda and {Gottl{\"o}ber}, Stefan and {Yepes}, Gustavo and {Libeskind}, Noam I. and {Pilipenko}, Sergey V. and {Knebe}, Alexander and {Courtois}, H{\'e}l{\`e}ne and {Tully}, R. Brent and {Steinmetz}, Matthias},
        title = "{Constrained Local UniversE Simulations: a Local Group factory}",
      journal = {\mnras},
         year = 2016,
        month = may,
       volume = {458},
       number = {1},
        pages = {900-911},
          doi = {10.1093/mnras/stw357},
archivePrefix = {arXiv},
       eprint = {1602.03919},
 primaryClass = {astro-ph.CO},
       adsurl = {https://ui.adsabs.harvard.edu/abs/2016MNRAS.458..900C}
}

@ARTICLE{Sorce2016,
       author = {{Sorce}, Jenny G. and {Gottl{\"o}ber}, Stefan and {Yepes}, Gustavo and {Hoffman}, Yehuda and {Courtois}, Helene M. and {Steinmetz}, Matthias and {Tully}, R. Brent and {Pomar{\`e}de}, Daniel and {Carlesi}, Edoardo},
        title = "{Cosmicflows Constrained Local UniversE Simulations}",
      journal = {\mnras},
         year = 2016,
        month = jan,
       volume = {455},
       number = {2},
        pages = {2078-2090},
          doi = {10.1093/mnras/stv2407},
archivePrefix = {arXiv},
       eprint = {1510.04900},
 primaryClass = {astro-ph.CO},
       adsurl = {https://ui.adsabs.harvard.edu/abs/2016MNRAS.455.2078S}
}

@ARTICLE{Rockstar2013,
       author = {{Behroozi}, Peter S. and {Wechsler}, Risa H. and {Wu}, Hao-Yi},
        title = "{The ROCKSTAR Phase-space Temporal Halo Finder and the Velocity Offsets of Cluster Cores}",
      journal = {\apj},
         year = 2013,
        month = jan,
       volume = {762},
       number = {2},
          eid = {109},
        pages = {109},
          doi = {10.1088/0004-637X/762/2/109},
archivePrefix = {arXiv},
       eprint = {1110.4372},
 primaryClass = {astro-ph.CO},
       adsurl = {https://ui.adsabs.harvard.edu/abs/2013ApJ...762..109B}
}

@article{
Muller2018,
author = {Oliver Müller  and Marcel S. Pawlowski  and Helmut Jerjen  and Federico Lelli },
title = {A whirling plane of satellite galaxies around Centaurus A challenges cold dark matter cosmology},
journal = {Science},
volume = {359},
number = {6375},
pages = {534-537},
year = {2018},
doi = {10.1126/science.aao1858},
URL = {https://www.science.org/doi/abs/10.1126/science.aao1858},
eprint = {https://www.science.org/doi/pdf/10.1126/science.aao1858}}

@ARTICLE{Tully2015,
       author = {{Tully}, R. Brent and {Libeskind}, Noam I. and {Karachentsev}, Igor D. and {Karachentseva}, Valentina E. and {Rizzi}, Luca and {Shaya}, Edward J.},
        title = "{Two Planes of Satellites in the Centaurus A Group}",
      journal = {\apjl},
         year = 2015,
        month = apr,
       volume = {802},
       number = {2},
          eid = {L25},
        pages = {L25},
          doi = {10.1088/2041-8205/802/2/L25},
archivePrefix = {arXiv},
       eprint = {1503.05599},
 primaryClass = {astro-ph.GA},
       adsurl = {https://ui.adsabs.harvard.edu/abs/2015ApJ...802L..25T}
}

@ARTICLE{Sawala2024,
       author = {{Sawala}, Till and {Frenk}, Carlos and {Jasche}, Jens and {Johansson}, Peter H. and {Lavaux}, Guilhem},
        title = "{Distinct distributions of elliptical and disk galaxies across the Local Supercluster as a {\ensuremath{\Lambda}}CDM prediction}",
      journal = {Nature Astronomy},
         year = 2024,
        month = feb,
       volume = {8},
        pages = {247-255},
          doi = {10.1038/s41550-023-02130-6},
archivePrefix = {arXiv},
       eprint = {2311.00755},
 primaryClass = {astro-ph.GA},
       adsurl = {https://ui.adsabs.harvard.edu/abs/2024NatAs...8..247S}
}

@ARTICLE{PlanckXVI,
       author = {{Planck Collaboration XVI}},
        title = "{Planck 2013 results. XVI. Cosmological parameters}",
      journal = {\aap},
         year = 2014,
        month = nov,
       volume = {571},
          eid = {A16},
        pages = {A16},
          doi = {10.1051/0004-6361/201321591},
archivePrefix = {arXiv},
       eprint = {1303.5076},
 primaryClass = {astro-ph.CO},
       adsurl = {https://ui.adsabs.harvard.edu/abs/2014A&A...571A..16P}
}

@ARTICLE{Kafle2018,
       author = {{Kafle}, Prajwal R. and {Sharma}, Sanjib and {Lewis}, Geraint F. and {Robotham}, Aaron S.~G. and {Driver}, Simon P.},
        title = "{The need for speed: escape velocity and dynamical mass measurements of the Andromeda galaxy}",
      journal = {\mnras},
         year = 2018,
        month = apr,
       volume = {475},
       number = {3},
        pages = {4043-4054},
          doi = {10.1093/mnras/sty082},
archivePrefix = {arXiv},
       eprint = {1801.03949},
 primaryClass = {astro-ph.GA},
       adsurl = {https://ui.adsabs.harvard.edu/abs/2018MNRAS.475.4043K}
}

@ARTICLE{Tamm2012,
       author = {{Tamm}, A. and {Tempel}, E. and {Tenjes}, P. and {Tihhonova}, O. and {Tuvikene}, T.},
        title = "{Stellar mass map and dark matter distribution in M 31}",
      journal = {\aap},
         year = 2012,
        month = oct,
       volume = {546},
          eid = {A4},
        pages = {A4},
          doi = {10.1051/0004-6361/201220065},
archivePrefix = {arXiv},
       eprint = {1208.5712},
 primaryClass = {astro-ph.CO},
       adsurl = {https://ui.adsabs.harvard.edu/abs/2012A&A...546A...4T}
}

@ARTICLE{Diaz2014,
       author = {{Diaz}, J.~D. and {Koposov}, S.~E. and {Irwin}, M. and {Belokurov}, V. and {Evans}, N.~W.},
        title = "{Balancing mass and momentum in the Local Group}",
      journal = {\mnras},
         year = 2014,
        month = sep,
       volume = {443},
       number = {2},
        pages = {1688-1703},
          doi = {10.1093/mnras/stu1210},
archivePrefix = {arXiv},
       eprint = {1405.3662},
 primaryClass = {astro-ph.GA},
       adsurl = {https://ui.adsabs.harvard.edu/abs/2014MNRAS.443.1688D}
}

@ARTICLE{Corbelli2010,
       author = {{Corbelli}, E. and {Lorenzoni}, S. and {Walterbos}, R. and {Braun}, R. and {Thilker}, D.},
        title = "{A wide-field H I mosaic of Messier 31. II. The disk warp, rotation, and the dark matter halo}",
      journal = {\aap},
         year = 2010,
        month = feb,
       volume = {511},
          eid = {A89},
        pages = {A89},
          doi = {10.1051/0004-6361/200913297},
archivePrefix = {arXiv},
       eprint = {0912.4133},
 primaryClass = {astro-ph.CO},
       adsurl = {https://ui.adsabs.harvard.edu/abs/2010A&A...511A..89C}
}

@ARTICLE{Posti2019,
       author = {{Posti}, Lorenzo and {Helmi}, Amina},
        title = "{Mass and shape of the Milky Way's dark matter halo with globular clusters from Gaia and Hubble}",
      journal = {\aap},
         year = 2019,
        month = jan,
       volume = {621},
          eid = {A56},
        pages = {A56},
          doi = {10.1051/0004-6361/201833355},
archivePrefix = {arXiv},
       eprint = {1805.01408},
 primaryClass = {astro-ph.GA},
       adsurl = {https://ui.adsabs.harvard.edu/abs/2019A&A...621A..56P}
}

@ARTICLE{Hattori2018,
       author = {{Hattori}, Kohei and {Valluri}, Monica and {Bell}, Eric F. and {Roederer}, Ian U.},
        title = "{Old, Metal-poor Extreme Velocity Stars in the Solar Neighborhood}",
      journal = {\apj},
         year = 2018,
        month = oct,
       volume = {866},
       number = {2},
          eid = {121},
        pages = {121},
          doi = {10.3847/1538-4357/aadee5},
archivePrefix = {arXiv},
       eprint = {1805.03194},
 primaryClass = {astro-ph.GA},
       adsurl = {https://ui.adsabs.harvard.edu/abs/2018ApJ...866..121H}
}

@ARTICLE{Monari2018,
       author = {{Monari}, G. and {Famaey}, B. and {Carrillo}, I. and {Piffl}, T. and {Steinmetz}, M. and {Wyse}, R.~F.~G. and {Anders}, F. and {Chiappini}, C. and {Jan{\ss}en}, K.},
        title = "{The escape speed curve of the Galaxy obtained from Gaia DR2 implies a heavy Milky Way}",
      journal = {\aap},
         year = 2018,
        month = aug,
       volume = {616},
          eid = {L9},
        pages = {L9},
          doi = {10.1051/0004-6361/201833748},
archivePrefix = {arXiv},
       eprint = {1807.04565},
 primaryClass = {astro-ph.GA},
       adsurl = {https://ui.adsabs.harvard.edu/abs/2018A&A...616L...9M}
}

@ARTICLE{Watkins2019,
       author = {{Watkins}, Laura L. and {van der Marel}, Roeland P. and {Sohn}, Sangmo Tony and {Evans}, N. Wyn},
        title = "{Evidence for an Intermediate-mass Milky Way from Gaia DR2 Halo Globular Cluster Motions}",
      journal = {\apj},
         year = 2019,
        month = mar,
       volume = {873},
       number = {2},
          eid = {118},
        pages = {118},
          doi = {10.3847/1538-4357/ab089f},
archivePrefix = {arXiv},
       eprint = {1804.11348},
 primaryClass = {astro-ph.GA},
       adsurl = {https://ui.adsabs.harvard.edu/abs/2019ApJ...873..118W}
}

@ARTICLE{Zaritsky2017,
       author = {{Zaritsky}, Dennis and {Courtois}, Helene},
        title = "{A dynamics-free lower bound on the mass of our Galaxy}",
      journal = {\mnras},
         year = 2017,
        month = mar,
       volume = {465},
       number = {3},
        pages = {3724-3728},
          doi = {10.1093/mnras/stw2922},
archivePrefix = {arXiv},
       eprint = {1611.04574},
 primaryClass = {astro-ph.GA},
       adsurl = {https://ui.adsabs.harvard.edu/abs/2017MNRAS.465.3724Z}
}

@article{Pierpaoli2003,
    author = {Pierpaoli, Elena and Borgani, Stefano and Scott, Douglas and White, Martin},
    title = {On determining the cluster abundance normalization},
    journal = {Monthly Notices of the Royal Astronomical Society},
    volume = {342},
    number = {1},
    pages = {163-175},
    year = {2003},
    month = {06},
    issn = {0035-8711},
    doi = {10.1046/j.1365-8711.2003.06525.x},
    url = {https://doi.org/10.1046/j.1365-8711.2003.06525.x},
    eprint = {https://academic.oup.com/mnras/article-pdf/342/1/163/3572367/342-1-163.pdf},
}

@ARTICLE{Valade2026,
       author = {{Valade}, Aur{\'e}lien and {Libeskind}, Noam and {Pomar{\`e}de}, Daniel and {Stiskalek}, Richard and {Hoffman}, Yehuda and {Gottl{\"o}ber}, Stefan and {Tully}, R. Brent},
        title = "{Constraining cosmological simulations with peculiar velocities: a forward-modeling approach}",
      journal = {arXiv e-prints},
         year = 2026,
        month = feb,
          eid = {arXiv:2602.03699},
        pages = {arXiv:2602.03699},
          doi = {10.48550/arXiv.2602.03699},
archivePrefix = {arXiv},
       eprint = {2602.03699},
 primaryClass = {astro-ph.CO},
       adsurl = {https://ui.adsabs.harvard.edu/abs/2026arXiv260203699V}
}

@ARTICLE{Ragagnin2021,
       author = {{Ragagnin}, Antonio and {Saro}, Alexandro and {Singh}, Priyanka and {Dolag}, Klaus},
        title = "{Cosmology dependence of halo masses and concentrations in hydrodynamic simulations}",
      journal = {\mnras},
         year = 2021,
        month = jan,
       volume = {500},
       number = {4},
        pages = {5056-5071},
          doi = {10.1093/mnras/staa3523},
archivePrefix = {arXiv},
       eprint = {2011.05345},
 primaryClass = {astro-ph.CO},
       adsurl = {https://ui.adsabs.harvard.edu/abs/2021MNRAS.500.5056R}
}

@ARTICLE{Diemer2015,
       author = {{Diemer}, Benedikt and {Kravtsov}, Andrey V.},
        title = "{A Universal Model for Halo Concentrations}",
      journal = {\apj},
         year = 2015,
        month = jan,
       volume = {799},
       number = {1},
          eid = {108},
        pages = {108},
          doi = {10.1088/0004-637X/799/1/108},
archivePrefix = {arXiv},
       eprint = {1407.4730},
 primaryClass = {astro-ph.CO},
       adsurl = {https://ui.adsabs.harvard.edu/abs/2015ApJ...799..108D}
}

@ARTICLE{Dutton2014,
       author = {{Dutton}, Aaron A. and {Macci{\`o}}, Andrea V.},
        title = "{Cold dark matter haloes in the Planck era: evolution of structural parameters for Einasto and NFW profiles}",
      journal = {\mnras},
         year = 2014,
        month = jul,
       volume = {441},
       number = {4},
        pages = {3359-3374},
          doi = {10.1093/mnras/stu742},
archivePrefix = {arXiv},
       eprint = {1402.7073},
 primaryClass = {astro-ph.CO},
       adsurl = {https://ui.adsabs.harvard.edu/abs/2014MNRAS.441.3359D}
}

\appendix

\section{$M_{200}/L_{K}$ conversion}
\label{sec:app1}

The conversion from luminosity to halo mass in the LVG catalog is limited by the amount of information contained in this survey, which forces us to assume certain general approximations. The main inconvenience arises from the fact that LVG contains luminosity in only two bands, B and K. This limits the empirical and semi-empirical $M_{\star}/L$ that can be used since only two independent variables are available. In this situation, infrared color-independent constant $M_{\star}/L_{K}$ have traditionally been used, the best-known case being $M_{\star}/L_{K} \sim 1$ from \citet{Bell2003}. However, this relation has the problem that it overestimates the stellar mass, so we have used other more recent estimates. 

The first, (L+RP), combines $M_{\star}/L_K$ expressed in the equation (\ref{eq:Lelli}) and a semi-empirical SHMR from \citet{Puebla2017}. \citet{McGaugh2014} shows that in order to improve the mass estimate using infrared bands and to make it self-consistent with the estimate of other bluer bands, it is required to assume the approximately constant ratio $\mathrm{log}(M_{\star}/L) = 0.47 M_{\odot}/L_{\odot}$ in $3.6 \mu m$. This is equivalent to the equation (\ref{eq:Lelli}) in the $K$ band. Later \citet{Lelli2016} confirms this correlation using rotation curves of nearby galaxies. Note that since the stellar mass of galaxies is determined by the different characteristics of the halo and its environment, the SHMR has a given scatter less than $\sigma_h \sim 0.15$ dex \citep{Puebla2015,Puebla2017}.

The second relation, (K+RP), uses $M_{\star}/L_K$ obtained by infrared SED estimation of the filter with Gaussian profile centered on the central wavelength of the band $K$ (also known as $K_s$), $\lambda_c = 2.16 \mu m$:
\begin{equation}
    \mathrm{log}(M_{\star}/L_{K}) = - 0.961+0.073 \, \mathrm{log}(M_{\star}/M_{\odot}) 
\label{eq:kim}
\end{equation}
with a scatter of 0.094 dex in $\mathrm{log}(M_{\star}/L)$. 

Finally, the third relation (KT) uses a direct relation between $L_{K}$ and $M_h$, which predicts predicts that the luminosity dependence of $\mathrm{log}(M_{h}/L_{K})$ decreases for low-mass galaxies in the range $\mathrm{log}(L_{K}/L_{\odot}) < 9$ and increases for groups and clusters above $\mathrm{log}(L_{K}/L_{\odot}) > 10.7$. Since we work with the individual luminosity of intermediate galaxies outside clusters, we will use the relation of \citet{Kourkchi2017}) corresponding to the first V-profile regime:
\begin{equation}
    \mathrm{log}(M_{h}/L_{K}) = 1.5 - 0.5\,\mathrm{log}(L_{K}/10^{10} L_{\odot})
\label{eq:deltah}
\end{equation}
This relation has been tested in LVG previously in satellite kinematics derivations of the halo mass \citep{Karachentsev2021}. All $\mathrm{log}(M_{h}/L_{K})$ ratios can be expressed in terms of $M_{200}$ by means of the relation $M_h \approx 1.22 M_{200}$ \citep{White2001,Pierpaoli2003}. 

\end{document}